\documentclass[journal,draftcls,onecolumn,12pt,twoside]{IEEEtranTCOM}
\normalsize

\usepackage{graphicx}
\usepackage{subfigure}
\usepackage{algorithm}
\usepackage{algcompatible}
\usepackage{amssymb}

\usepackage{color,xcolor}
\usepackage{bm}
\usepackage[multiple]{footmisc}
\usepackage{booktabs}

\usepackage{cite}

\usepackage{mathrsfs}

\ifCLASSINFOpdf
\else
\fi

\usepackage[cmex10]{amsmath}
\usepackage{array}

\usepackage{mdwmath}
\usepackage{mdwtab}

\usepackage{eqparbox}
\definecolor{red}{rgb}{1,0,0}
\usepackage{color,soul}
\definecolor{lightblue}{rgb}{.90,.95,1}
\sethlcolor{yellow} 

\newcommand{\myBlue}[1]{\textcolor{black}{#1}}

\begin{document}
\font\myfont=cmr12 at 22pt

\title{\myfont \myBlue{Error Propagation Mitigation in Sliding Window Decoding} of Braided Convolutional Codes}
%
%
%

\author{\normalsize{Min~Zhu,~\IEEEmembership{\normalsize{Member,~IEEE,}}
        David~G.~M.~Mitchell,~\IEEEmembership{\normalsize{Senior Member,~IEEE,}}
        Michael~Lentmaier,~\IEEEmembership{\normalsize{Senior Member,~IEEE,}\\}
        Daniel J. Costello, Jr.,~\IEEEmembership{\normalsize{Life~Fellow,~IEEE,}}
        and~Baoming~Bai,~\IEEEmembership{\normalsize{Senior Member,~IEEE}}}
\thanks{This material is based upon work supported by the National Science Foundation under Grant Nos. ECCS-1710920 and OIA-1757207. This work was presented in part at the Information Theory and Applications Workshop, San Diego, CA, February 2018, and at the International Symposium on Information Theory, Vail, Colarado, USA, June 2018.}
\thanks{M.~Zhu  and B. Bai are with the State Key Lab.~of ISN, Xidian University, Xi'an 710071, China,
(e-mail:~zhunanzhumin@gmail.com;~bmbai@mail.xidian.edu.cn).}
\thanks{D.~G.~M.~Mitchell is with the Klipsch School of Electrical and Computer Engineering, New Mexico State University, Las Cruces, NM 88003, USA, (e-mail:~dgmm@nmsu.edu).}
\thanks{M. Lentmaier is with the Department of Electrical and Information Technology, Lund University, 221 00 Lund, Sweden. (e-mail:~michael.lentmaier@eit.lth.se).}
\thanks{D.~J. Costello, Jr. is with the Department of Electrical Engineering, University of Notre Dame, Notre Dame, IN 46556, USA, (e-mail:~dcostel1@nd.edu).}}

%
%

\markboth{IEEE Transactions on Communications}%
{Submitted paper}
%



\maketitle

\begin{abstract}
\myBlue{We investigate error propagation in sliding window decoding of braided convolutional codes (BCCs).} Previous studies of BCCs have focused on iterative decoding thresholds, minimum distance properties, and \myBlue{their} bit error rate (BER) performance \myBlue{at small to moderate frame length. Here, we consider a sliding window decoder in the context of large frame length or one that} continuously outputs blocks in a streaming fashion. In this case, decoder error propagation, due to the feedback inherent in BCCs, can be a serious problem.
In order to mitigate the effects of error propagation, we propose several schemes: a \emph{window extension algorithm} where the decoder window size can be extended adaptively, a \emph{resynchronization mechanism} where we reset the encoder to the initial state, and a \emph{retransmission strategy} where \myBlue{erroneously decoded} blocks are retransmitted. In addition, we introduce a soft BER stopping rule to reduce computational complexity, and the tradeoff between performance and complexity is examined. Simulation results show that, using the proposed window extension algorithm, resynchronization mechanism, and retransmission strategy, the BER performance of BCCs can be improved by up to four orders of magnitude in the signal-to-noise ratio operating range of interest, and in addition the soft BER stopping rule can be employed to reduce computational complexity.
\end{abstract}


%
\IEEEpeerreviewmaketitle

\section{Introduction}
%
%
%
%
\IEEEPARstart{B}{raided} convolutional codes (BCCs), first introduced in \cite{WeizhangIT2010}, are a counterpart to braided block codes (BBCs)~\cite{FeltstromIT2009},\footnote{A type of BBC, braided Bose-Chaudhuri-Hocqenghem (BCH) codes \cite{Jian2013GLOBECOM}, and the closely related staircase codes \cite{Benjamin2012,Amat2018} have been investigated for high speed optical communication.} which can be regarded as a diagonalized version of product codes \cite{Elias1954} or expander codes \cite{Sipser1996}. In contrast to BBCs, BCCs use short constraint length convolutional codes as component codes. The encoding of BCCs can be described by a two-dimensional sliding array of encoded symbols, where each symbol is protected by two component convolutional codes. In this sense, BCCs are a type of parallel-concatenated (turbo) code in which the parity outputs of one component encoder are fed back and used as inputs to the other component encoder at the succeeding time unit. Two variants of BCCs, tightly and sparsely braided codes, were considered in \cite{WeizhangIT2010}. Tightly braided convolutional codes (TBCCs) are obtained if a dense array is used to store the information and parity symbols. This construction is deterministic and simple to implement but performs relatively poorly due to the absence of randomness. Alternatively, sparsely braided convolutional codes (SBCCs) employ random permutors and have ``turbo-like'' code properties, resulting in improved iterative decoding performance \cite{WeizhangIT2010}. SBCCs can operate in either a bitwise or blockwise mode, depending on whether convolutional or block permutors are employed. Moloudi \textit{et al}. characterized SBCCs as a type of spatially coupled turbo code with a regular graph structure and showed that threshold saturation occurs for iterative decoding of SBCCs over the binary erasure channel \cite{Saeedeh2017IT,Saeedeh2016ISITA}, and Farooq \textit{et al}. proposed a technique to compute the thresholds of SBCCs on the additive white Gaussian noise (AWGN) channel \cite{Farooq2018ISIT}. It was also shown numerically that the free (minimum) distance of bitwise and blockwise SBCCs grows linearly with the overall constraint length, leading to the conjecture that SBCCs, unlike parallel or serial concatenated codes, are asymptotically good \cite{WeizhangIT2010,Saeedeh2016ISITA,Saeedeh2019TCOM}.

Due to their turbo-like structure, SBCCs can be decoded with iterative decoding. Analogous to LDPC convolutional codes \cite{LentmaierTIT2010, IyengarTIT2012}, SBCCs can employ sliding window decoding \myBlue{(SWD)} for low latency operation \cite{Zhu2017TCOM}. Unlike \myBlue{SWD} of LDPC convolutional codes, which typically uses an iterative belief-propagation (BP) message passing algorithm, \myBlue{SWD} of SBCCs is based on the Bahl-Cocke-Jelinek-Raviv (BCJR) algorithm. It has been shown that blockwise SBCCs with \myBlue{SWD} have excellent performance \cite{Zhu2017TCOM}, but for large frame lengths or streaming (continuous transmission) applications, it has been observed that SBCCs are susceptible to infrequent but severe decoder error propagation \cite{Zhu2018ISIT}. That is, once a block decoding error occurs, decoding of the following blocks can be affected, which in turn can cause a continuous string of block errors and result in unacceptable performance loss. \myBlue{Although streaming codes have been widely investigated \cite{Kuma2020TIT,Badr2017TIT,Badr2015TIT,Dudzicz2020TCOM}, our paper focuses only on the use of capacity-approaching codes and the desire to limit latency in such cases by employing SWD. To our knowledge, the only other work to consider the error propagation problem with SWD of capacity-approaching codes is the recent paper by Klaiber et al. (\hspace{1sp}\cite{Brink2018ISTC}). That paper considered \emph{spatially coupled LDPC codes} and the mitigation methods developed there, including adapting the number of iterations and window shifting, are different from the ones we propose for BCCs.}

In this paper, we examine the causes of error propagation in \myBlue{SWD} of SBCCs and propose several error propagation mitigation techniques. Specifically, based on a prediction of the reliability of a decoded block, a \emph{window extension algorithm}, a \emph{resynchronization mechanism}, and a \emph{retransmission strategy} are introduced to combat the error propagation. In addition, a soft bit-error-rate stopping rule is proposed to reduce decoding complexity and the resulting tradeoff between decoding performance and decoding complexity is explored.

\section{\myBlue{Review of Braided Convolutional Codes}}
In this section, we briefly review \myBlue{the encoding and SWD} of blockwise SBCCs. For further details, please refer to \cite{WeizhangIT2010} and \cite{Zhu2017TCOM}.

\subsection{\myBlue{Encoding}}
\begin{figure}\center
        \includegraphics[width= 0.6 \textwidth]{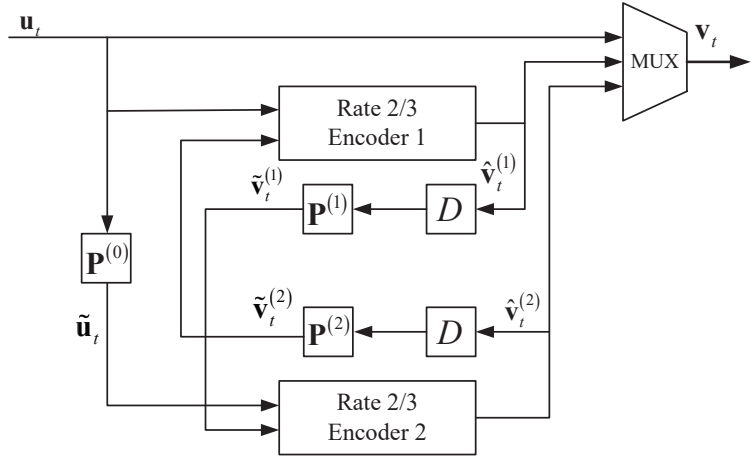}\\
        \vspace{-5mm}
        \caption{Encoder for a rate $R = 1/3$ blockwise SBCC.}
        \label{fig:Encoder}
    \end{figure}

\begin{figure*}[t]
        \center
        \includegraphics[width= \textwidth]{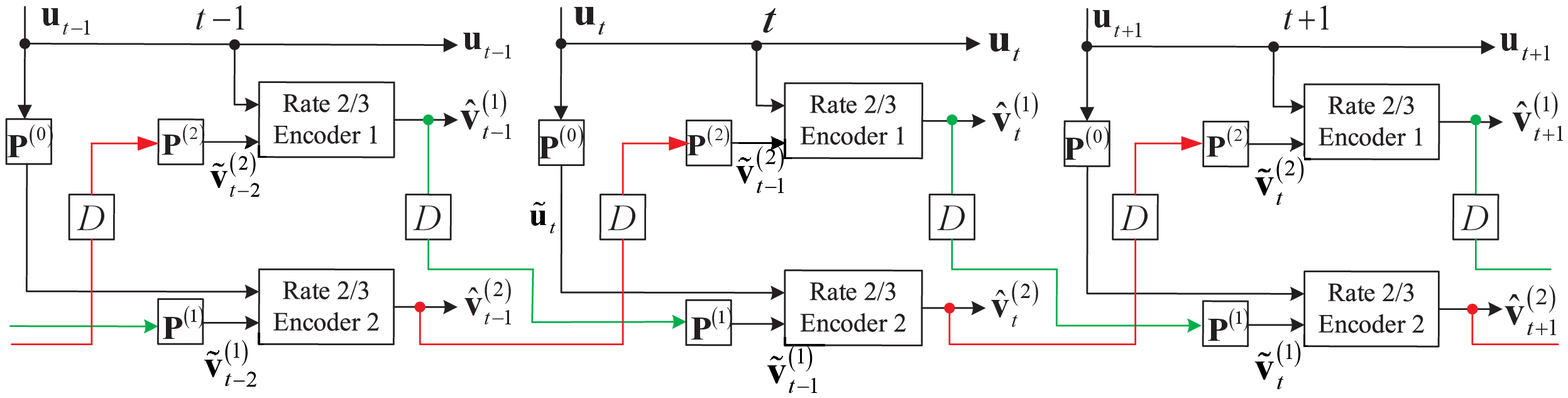}\\
        \vspace{-5mm}
        \caption{\myBlue{Encoder} chain for a rate $R = 1/3$ blockwise SBCC.}
        \label{fig:Chain_Encoder}
        \vspace{-2mm}
    \end{figure*}

SBCCs are constructed using a turbo-like parallel concatenation of two component encoders. However, unlike turbo codes, the two encoders share parity feedback. In this manner, the systematic and parity symbols are ``braided'' together. In this paper, we restrict our discussion to rate $R = 1/3$ blockwise SBCCs, but generalization to other rates and to bitwise SBCCs is straightforward. In this case, the information sequence enters the encoder in a block-by-block manner, typically with a relatively large block size. Fig. \ref{fig:Encoder} shows the \myBlue{encoding} process for a rate $R = 1/3$ blockwise SBCC, which utilizes two recursive systematic convolutional (RSC) component encoders each of rate ${R_{cc}} = 2/3$, where ${{\bf{P}}^{\left( 0 \right)}}$, ${{\bf{P}}^{\left( 1 \right)}}$, and ${{\bf{P}}^{\left( 2 \right)}}$ are each block permutors of length $T$. The information sequence is divided into blocks of length $T$ symbols, i.e., ${\bf{u}} = \left( {{{\bf{u}}_0},{{\bf{u}}_1}, \ldots ,{{\bf{u}}_t}, \ldots } \right)$, where ${{\bf{u}}_t} = \left( {{u_{t,1}},{u_{t,2}}, \ldots ,{u_{t,T}}} \right)$. At time $t$, ${{\bf{u}}_t}$ is interleaved using ${{\bf{P}}^{\left( 0 \right)}}$ to form ${{\bf{\tilde u}}_t}$, and ${{\bf{u}}_t}$ and ${{\bf{\tilde u}}_t}$ enter the component encoders. The parity outputs $\hat {\bf{v}}_{t}^{\left( i \right)}$ from encoder $i$, $i \in \{ 1,2\} $, at time $t$ are delayed by one time unit, interleaved using ${{\bf{P}}^{\left( 1 \right)}}$ and ${{\bf{P}}^{\left( 2 \right)}}$, respectively, and then enter the component encoders as the input sequences $\tilde {\bf{v}}_{ t+1}^{\left( i \right)}$, $i \in \{ 1,2\} $, at time $t+1$. The information block ${\bf{u}}_t$, the parity output block ${\bf{\hat v}}_t^{\left( 1 \right)}$ of encoder 1, and the parity output block ${\bf{\hat v}}_t^{\left( 2 \right)}$ of encoder 2 are sent over the channel as the encoded block ${\bf{v}}_t = \left( {\bf{u}}_t, {\bf{\hat v}}_t^{\left( 1 \right)}, {\bf{\hat v}}_t^{\left( 2 \right)}\right)$ at time $t$.
In order to depict the \myBlue{encoding} process conceptually, a chain of encoders that operates at different time instants is illustrated in Fig. \ref{fig:Chain_Encoder}. At each time instant, there is a turbo-like encoder which consists of two parallel concatenated RSC component encoders. These turbo-like encoders are coupled by feeding the parity sequence generated at the current time instant to the encoders at the next time instant, so that the coupling memory is 1 in this case.
For initialization, at time instant 0, we assume that ${\bf{\tilde v}}_{ - 1}^{\left( 1 \right)} = {\bf{0}}$ and ${\bf{\tilde v}}_{ - 1}^{\left( 2 \right)} = {\bf{0}}$.

Transmission can be \emph{terminated} after a \emph{frame} consisting of $L$ encoded blocks by inserting a small number $l$ of additional blocks (typically $l \ll L$), in which case the rate is given by $R_{L} = \frac{{\rm{1}}}{{\rm{3}}} \cdot \frac{L}{{L - l}}$ and we suffer a slight rate loss, or \emph{unterminated} (in a \myBlue{continuous} streaming fashion), in which case the rate is given by $ R = \frac{1}{3}$.


\subsection{Sliding Window Decoding}

\begin{figure*}[t]
        \center
        \includegraphics[width= \textwidth]{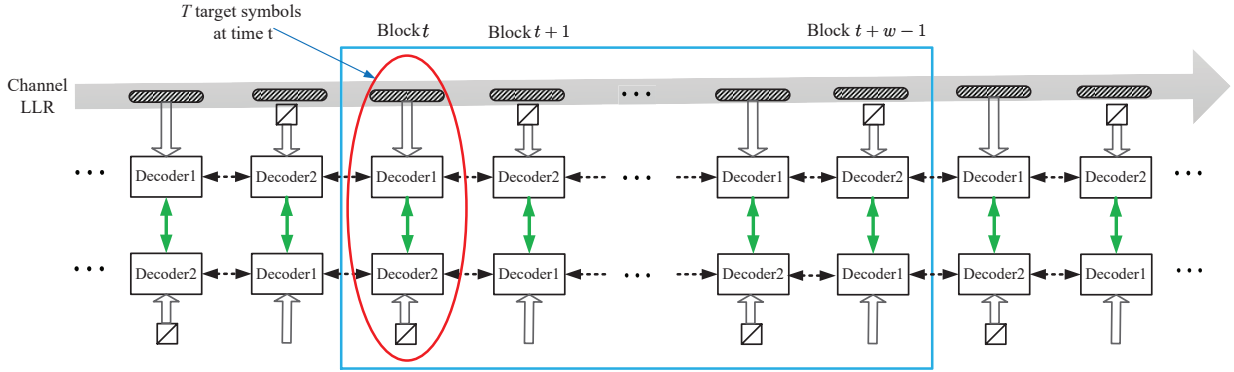}\\
        \vspace{-5mm}
        \caption{\myBlue{Sliding} window decoder for blockwise SBCCs \cite{Zhu2017TCOM}.}
        \label{fig:Decoder}
        \vspace{-2mm}
    \end{figure*}

In order to help describe the proposed error propagation mitigation methods, the structure of the sliding window decoder \cite{Zhu2017TCOM} is shown in Fig. \ref{fig:Decoder}. The \emph{window size} is denoted as $w $. The block at time instant $t$ is the \emph{target block} for decoding in the window containing the blocks received at times $t$ to $t+w-1$. The decoding process in a window beings with $I_1$ turbo, or \emph{vertical}, iterations on the target block at time $t$, during which the two component convolutional codes pass soft messages on the $T$ information bits in that block to each other. Then, soft messages on the parity bits are passed forward, and $I_1$ vertical iterations are performed on the block at time $t+1$. This continues until $I_1$ vertical iterations are performed on the last received block in the window. Then the process is repeated in the backward direction (from the last block to the first block in the window) with soft messages being passed back through the 2$w$ BCJR decoders. This round trip of decoding is called a \emph{horizontal} iteration. After $I_2$ horizontal iterations, the $T$ target symbols are decoded, and the window shifts forward to the next position, where the $T$ symbols at time $t+1$ become the target symbols.\footnote{Other decoding schedules were proposed in \cite{Zhu2017TCOM}, but those do not affect the general discussion in this paper.}

\section{Error Propagation}
Since an encoded block in a blockwise SBCC affects the encoding of the next block (see Fig. \ref{fig:Chain_Encoder}), each time a block of target symbols is decoded, the log-likelihood ratios (LLRs) associated with the decoded symbols also affect the decoding of the next block. Hence if, after a fixed maximum number of decoding iterations, some unreliable LLRs remain in the target block \myBlue{and} cause a block decoding error, those unreliable LLRs can potentially trigger a string of additional block errors, resulting in \emph{error propagation}.

\subsection{\myBlue{Motivation}}
\begin{figure}[h]
        \center
        \includegraphics[width= 0.65 \textwidth]{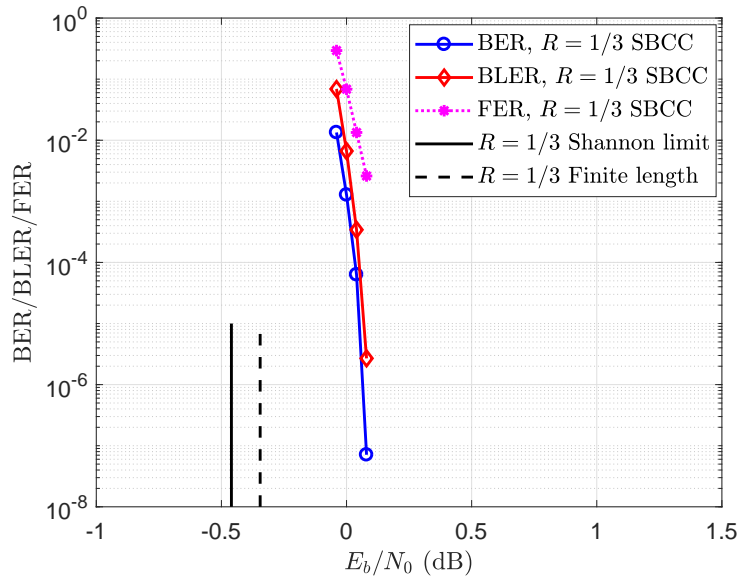}\\
        \vspace{-5mm}
        \caption{The BER, BLER, and FER performance of a rate $R=1/3$ SBCC with $T=8000$ and $L=1000$.}
        \label{fig:Origin}
        \vspace{-4mm}
    \end{figure}

\emph{Example 1}: To illustrate this effect, we consider an example of two identical 4-state RSC component encoders whose generator matrix is given by
\begin{equation}
\label{GenerateMatrix}
{G}\left( D \right) = \left( {\begin{array}{*{20}{c}}
   1 & 0 & {\frac{1}{{1 + D + {D^2}}}}  \\
   0 & 1 & {\frac{{1 + {D^2}}}{{1 + D + {D^2}}}}  \\
\end{array}} \right),
\end{equation}
where we assume the encoders are left unterminated at the end of each block. The three block permutors ${{\bf{P}}^{\left( 0 \right)}}$, ${{\bf{P}}^{\left( 1 \right)}}$, and ${{\bf{P}}^{\left( 2 \right)}}$ are assumed to be chosen randomly with the same size $T=8000$, and we also assume that transmission stops after a frame of $L$ blocks is decoded and a uniform decoding schedule is used (see \cite{Zhu2017TCOM} for details).\footnote{In a uniform decoding schedule, the same number of vertical iterations in the forward message passing process (from the first block to the last block in the decoding window) are performed on each block. Likewise, in the backward message passing process  (from the last block to the first block in the window), the same number of vertical iterations are performed.} The bit error rate (BER), block error rate (BLER), and frame error rate (FER) performance for transmission over the AWGN channel with BPSK signalling are plotted in Fig. \ref{fig:Origin} as functions of the \myBlue{channel} signal-to-noise ratio (SNR) $E_b/N_0$, where the window size $w=3$, the number of vertical iterations is $I_1 = 1$, the number of horizontal iteration is $I_2 = 20$, and the frame length is $L=1000$ blocks.

\begin{figure}[htbp]
  \setlength{\subfigcapskip}{-0.5cm}
   \centerline{\subfigure[]{\includegraphics[width=0.45 \textwidth]{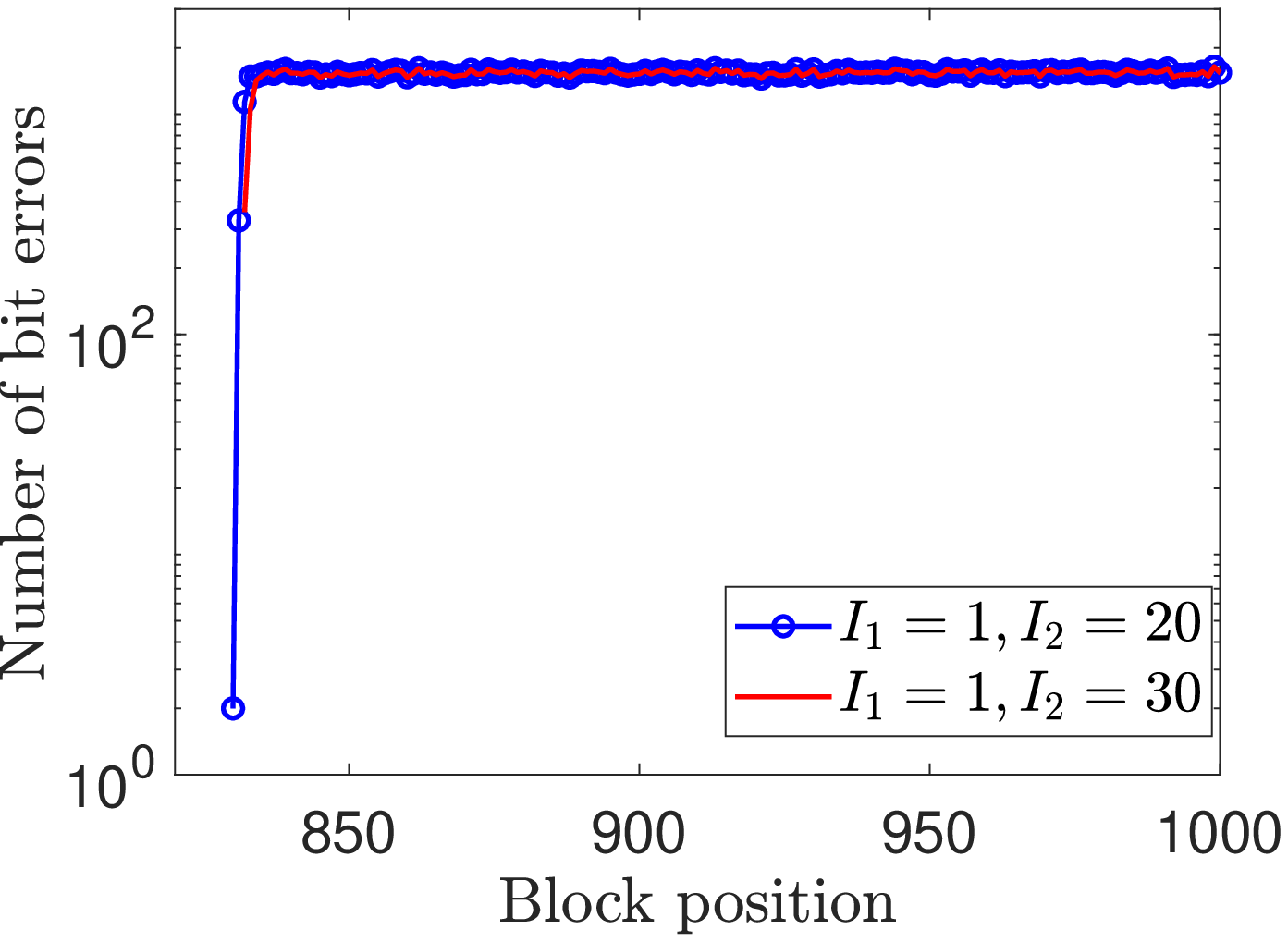}
       \label{fig:DiffITER}}
     \hfil
     \subfigure[]{\includegraphics[width=0.45 \textwidth]{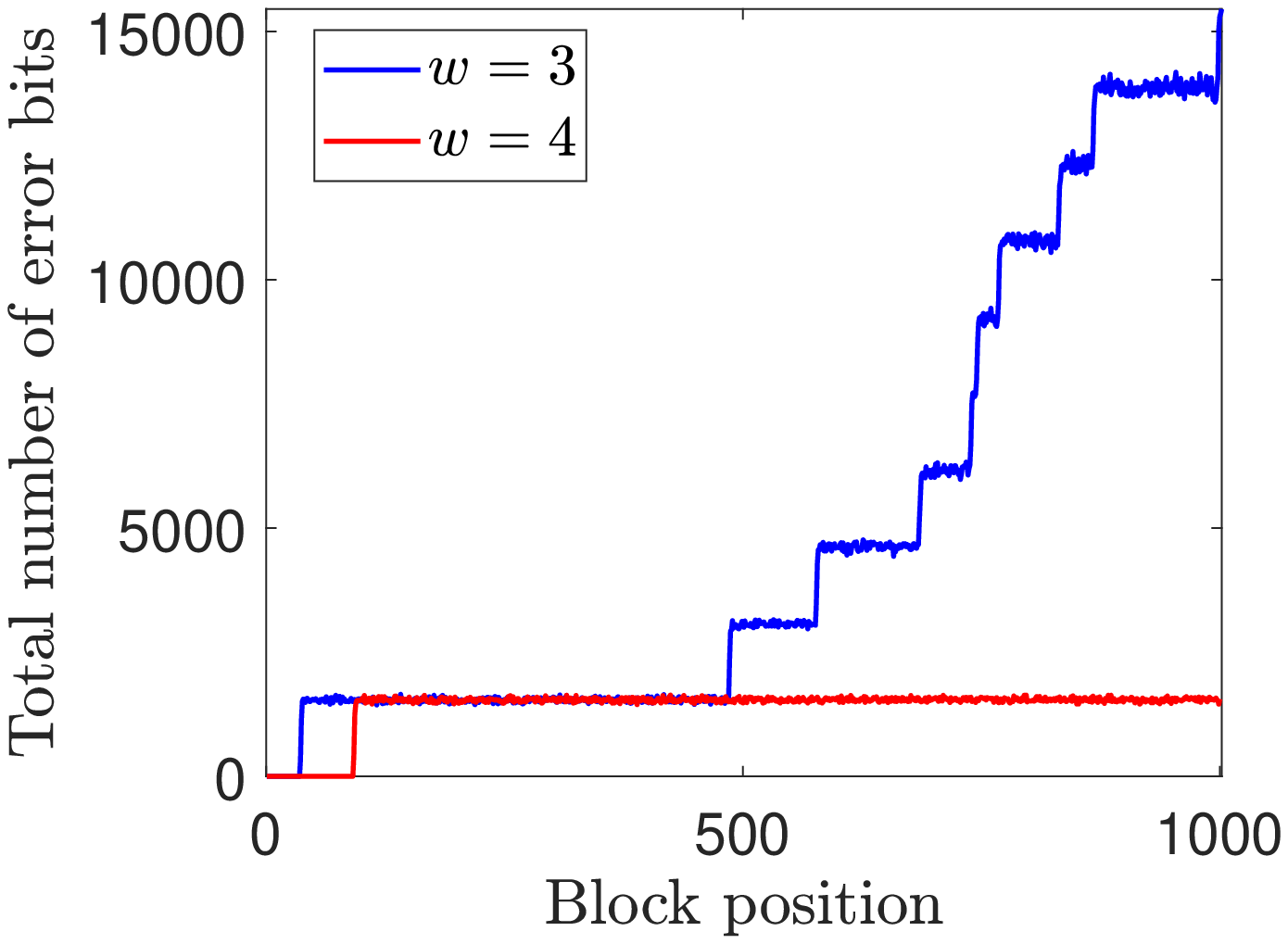}
       \label{fig:DiffWin}}}
   \setlength{\abovecaptionskip}{-0.01cm} 
   \caption{The bit error distribution per block for a rate $R=1/3$ blockwise SBCC with $T=8000$: (a) one frame with different numbers of iterations, $w=3$, and (b) 10000 frames with different window sizes, $I_1=1$, $I_1=20$.}
   \label{fig:ErrDis}
   \vspace{-3mm}
 \end{figure}

From Fig. \ref{fig:Origin}, we see that the rate $R=1/3$ blockwise SBCC performs about 0.5 dB away from the Shannon limit and 0.4 dB away from the finite-length bound\myBlue{\cite{PolyanskiyTIT2010}} at a BER of $10^{-6}$. Even so, among the 10000 simulated frames, several were observed to exhibit error propagation.  For example, 9 such frames were observed at $E_b/N_0 = 0.04$ dB. In order to depict the error propagation phenomenon clearly, we show the bit error distribution per block of one frame with error propagation in Fig. \ref{fig:DiffITER}. We see that, for $I_2=20$, from block 830 on, the number of error bits is large, and the errors continue to the end of the frame, a clear case of error propagation. For $I_2=30$, error propagation starts two blocks later than for $I_2=20$, but we see that the overall effect of increasing the number of iterations is minimal.\footnote{In related work on spatially coupled LDPC codes with sliding window decoding, Klaiber \emph{et al}.\myBlue{\cite{Brink2018ISTC}} have also noted a problem with error propagation and have successfully employed an adaptive number of iterations and window shifting to improve performance in that case.} On the other hand, the bit error distribution per block, based on 10000 simulated frames with two different window sizes, is shown in Fig. \ref{fig:DiffWin}, where we see that increasing the window size from 3 to 4 reduces the number of error propagation frames from 9 to 1, thus significantly improving performance.
\hfill $\blacksquare$

\subsection{\myBlue{A Decoder Model of Error Propagation}}
\myBlue{Assuming that information is transmitted in frames of length $L$, a significant number of blocks could be affected by error propagation if $L$ is large, thus severely degrading the BLER performance. We now give a brief analysis of how error propagation affects the BLER performance of SWD.}\footnote{\myBlue{A similar analysis was presented in a recent paper \cite{Zhu2020ISIT} on SWD of spatially coupled low-density parity-check (LDPC) codes.}}

\begin{figure}[htbp]
   \centering
   \includegraphics[width=0.3\textwidth]{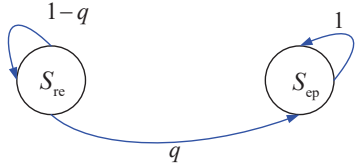}
   \caption{\myBlue{The state diagram describing the operation of a decoder subject to error propagation.}}
   \label{fig:StateDiagram}
   \vspace{-1mm}
 \end{figure}

\myBlue{Assume that, in any given frame, the decoder operates in one of two states: $(1)$ a \emph{random error state} ${S_{\rm{re}}}$ in which block errors occur independently with probability $p$, and $(2)$ an \emph{error propagation state} ${S_{\rm{ep}}}$ in which block errors occur with probability 1. Also assume that, at each time unit $t=1, 2, 3, \ldots,L$ the decoder transitions from state ${S_{\rm{re}}}$ to state ${S_{\rm{ep}}}$ independently with probability $q$ (typically, $q \ll p$) and that, once in state ${S_{\rm{ep}}}$, the decoder remains there for the rest of the frame.\footnote{A given frame can (1) operate entirely in state ${S_{\rm{re}}}$, where error propagation never occurs, (2) start in state ${S_{\rm{re}}}$ and then at some time transition to state ${S_{\rm{ep}}}$, or (3) operate entirely in state ${S_{\rm{ep}}}$, where the very first block is decoded incorrectly and block errors continue throughout the rest of the frame.} A state diagram describing this situation is shown in Fig. \ref{fig:StateDiagram}.}

\myBlue{Consider a simulation scenario in which the information block size $T$ and $B$, the total number of blocks to be simulated, are fixed, where $N$ is the total number of simulated frames, $B = LN$, and $TB$ is the total number of simulated symbols. Under normal (random error) decoder operating conditions, the simulated BLER should be independent of the particular combination of $L$ and $N$ chosen. When decoder error propagation is possible, however, we now show that, for fixed $B$, the values of $L$ and $N$ can affect the simulated BLER.}

\myBlue{For a frame of length $L$, we express the probability that the decoder first enters state $S_{\rm{ep}}$ at time $t = \tau$ (and thus stays in state $S_{\rm{ep}}$ until time $t=L$) as
 \begin{equation}
{P_\tau }\left( {{S_{\rm{ep}}},t = \left[ {\tau :L} \right]} \right) = q{\left( {1 - q} \right)^{\tau  - 1}},~~~{\rm{   }}\tau  = 1,2, \ldots ,L,
 \end{equation}
where the notation $t = \left[ {{t_1}:{t_2}} \right]$ denotes the set of time units from $t_1$ to $t_2$.}
\myBlue{Similarly, we can write the probability that the decoder stays in state $S_{\rm{re}}$ throughout the entire frame as
\begin{equation}
\begin{aligned}
P\left( {{S_{\rm{re}}},t = \left[ {1:L} \right]} \right) = 1 - \sum\limits_{\tau  = 1}^{L} {{P_\tau }\left( {{S_{\rm{ep}}},t = \left[ {\tau :L} \right]} \right)}= \left( {1 - q} \right)^{L}.
\end{aligned}
\end{equation}
Now, given that a frame enters state $S_{\rm{ep}}$ at time $t=\tau$, we can express the average BLER as
\begin{subequations}
\begin{equation}
{P_{\rm{BL}}}\left( {\tau  = 1} \right) = 1,\\
\end{equation}
\begin{equation}
\begin{aligned}
 {P_{\rm{BL}}}\left( \tau  \right) = {{\left[ {p \cdot \left( {\tau  - 2} \right) + L - \tau  + 1} \right]} \mathord{\left/
 {\vphantom {{\left[ {p \cdot \left( {\tau  - 2} \right) + L - \tau  + 1} \right]} L}} \right.
 \kern-\nulldelimiterspace} L},~~~\tau  = 2, \ldots ,L, \\
 \end{aligned}
\end{equation}
\label{eq:Pbl_Tau}%
\end{subequations}
where we note that state $S_{\rm{ep}}$ must be preceded by at least one correctly decoded block.  Finally, we can write the overall average BLER as
\begin{equation}
\begin{aligned}
P_{\rm{BL}} = \sum\limits_{\tau  = 1}^{L} {{P_{\rm{BL}}}\left( \tau  \right) \cdot q{\left( {1 - q} \right)^{\tau  - 1}}} + p \cdot \left( {1 - q} \right)^{L}.
\end{aligned}
\label{eq:Pbl}
\end{equation}}

\myBlue{Looking at \eqref{eq:Pbl}, it is clear that, if $q=0$, i.e., we never enter state $S_{\rm{ep}}$, then $P_{\rm{BL}}=p$, independent of the frame length $L$. This is the normal condition under which Monte Carlo simulations are conducted. However, under error propagation conditions, the simulated BLER will increase as a function of the frame length. We also note that the model parameters $p$ and $q$ will depend both on the channel SNR and the decoder window size $w$. In general, both lower SNRs and smaller values of $w$ will result in larger values of $p$ (random block error probability) and $q$ (error propagation probability), making the performance more sensitive to large values of $L$. By contrast, high SNR and large $w$ will reduce $p$ and $q$, making performance less sensitive to the value of $L$.}

\begin{figure}[t]
    \center
    \includegraphics[width= .65\textwidth]{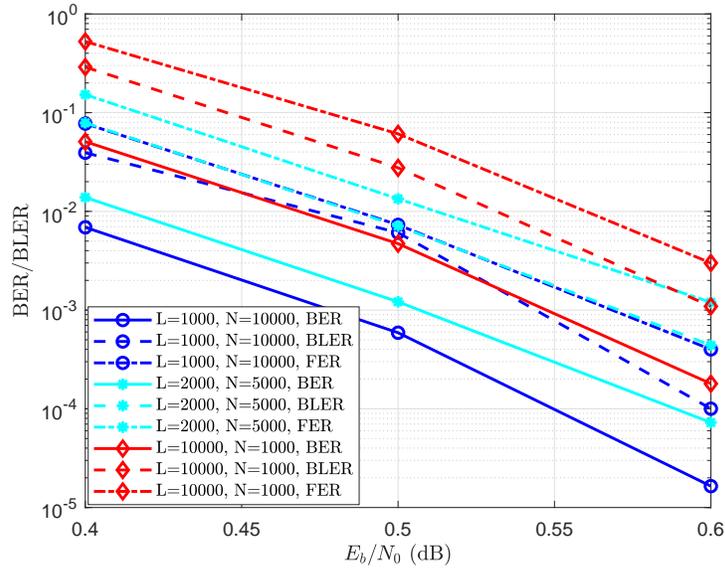}
    \caption{\myBlue{The BER, BLER, and FER performance of a $R=1/3$ BCC with $T$=1000, $w$=5, and different values of $L$ and $N$ such that $TLN = 10^{10}$.}}
    \label{fig:WinExtT1000DiffNL}       
\end{figure}

\myBlue{\emph{Example 2}: Consider a rate $R=1/2$ blockwise BCC with information block size $T=1000$, window size $w=5$, and different frame lengths $L$ and numbers of simulated frames $N$ such that the total number of simulated blocks is $B=LN=10^{7}$, or $TLN = 10^{10}$ simulated symbols. The BER, BLER, and FER performance is shown in Fig. \ref{fig:WinExtT1000DiffNL}.}

\myBlue{From Fig. \ref{fig:WinExtT1000DiffNL}, we see that the simulation runs with a larger frame length $L$ and a smaller number of simulated frames $N$ exhibit \emph{higher errors rates} than those with smaller $L$ and larger $N$, for the same total number of simulated blocks $B=LN$. Also note that, in a true streaming environment ($L \to \infty $), the BER will tend to 0.5 and both the BLER and FER will tend to 1.0!}
\hfill $\blacksquare$

\myBlue{This example makes clear that, for $L \gg 1000$, and particularly for streaming transmission, error propagation will severely degrade the decoding performance illustrated in Fig. \ref{fig:Origin}. In the next section, we look more carefully at the error propagation statistics, and then in Section V we introduce three ways of mitigating error propagation in sliding window decoding of SBCCs.}

\section{What Causes Error Propagation?}
In this section, we investigate the causes of error propagation during sliding window decoding of SBCCs. To this end, we introduce the concept of a \emph{superstate} to describe the complete state of the encoder at a given time $t$, i.e., the information needed to generate the $3T$-symbol encoded block $\textbf{v}_t$ from the $T$-symbol information block $\textbf{u}_t$. From Fig. \ref{fig:Encoder}, with the 4-state RSC component encoders of (\ref{GenerateMatrix}), we see that the superstate consists of the two T-bit parity input sequences from the previous block plus the four component encoder register bits at the beginning of a block, which together determine the output block $\textbf{v}_t$ for a given input block $\textbf{u}_t$.\myBlue{\footnote{The component RSC encoders are not terminated at the end of a block, so the register bits at the beginning of block $t$ are the same as those at the end of block $t-1$.}}

In the following, we give two examples with different permutor sizes to illustrate the causes of error propagation.

\begin{figure}
    \centering

    \includegraphics[width= 0.65 \textwidth]{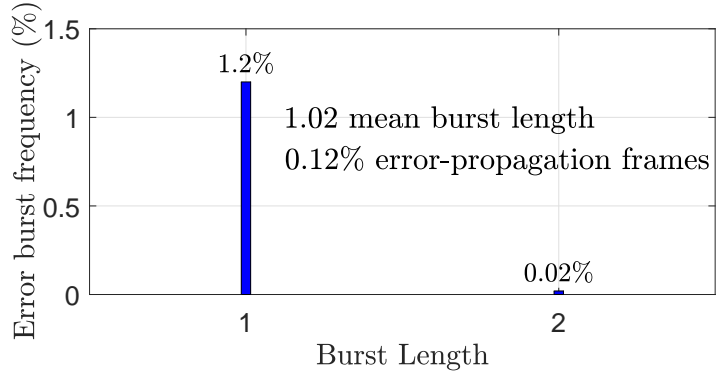}
  \setlength{\abovecaptionskip}{-0.3cm} 
  \caption{The frequency of the error frames in a rate $R=1/3$ SBCC with $T=8000$, $w=3$, $I_1=1, I_2=20$, and $L=1000$ at $E_b/N_0 = 0.04$ dB.}
  \label{fig:BurstDistWinP8000}
  \vspace{-2mm}
  \end{figure}

\myBlue{\emph{Example 3}}: We first consider the case of large permutor (block) size $T$.  10000 frames of the rate $R=1/3$ blockwise SBCC from Example 1 were simulated at $E_b/N_0 = 0.04$ dB (corresponding to a BER of about $10^{-7}$) with $T=8000$, $w=3$, $I_1 = 1$, $I_2 = 20$, and $L=1000$. LLRs are capped at $\pm 20$. The simulated frames consisted of correct frames, frames with short bursts of one or two block errors, and error-propagation \myBlue{frames.\footnote{
When consecutive error  blocks continue to the end of a frame, we call it \emph{error propagation}. When the last block in a sequence of one or more consecutive error  blocks does not coincide with the end of a frame, we call it a \myBlue{\emph{burst error}.}}} The frequency of the burst-error frames and error-propagation frames among the 10000 simulated frames, \myBlue{along with the mean burst length,} is shown in \myBlue{Fig. \ref{fig:BurstDistWinP8000}.\footnote{Note that, since there may be multiple burst errors in a frame, or a frame may contain burst errors along with error propagation, the total number of burst errors may exceed the number of frames containing burst errors.}}

\begin{figure}
        \center
        \includegraphics[width= 0.65 \textwidth]{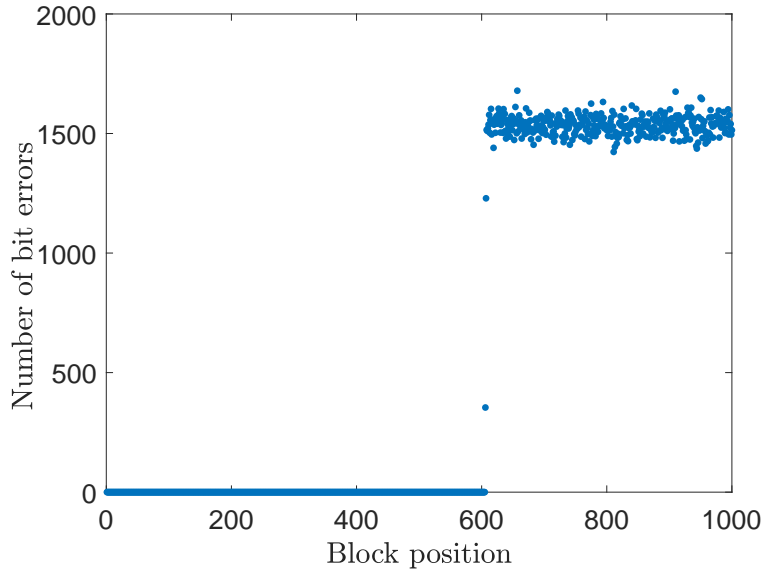}\\
        \setlength{\abovecaptionskip}{-0.3cm} 
        \caption{The bit error distribution per block in an example error-propagation frame from a rate $R=1/3$ SBCC with $T=8000$.}
        \label{fig:ErrDist4917}
        \vspace{-7mm}
\end{figure}

\myBlue{Fig. \ref{fig:ErrDist4917}} shows the bit error distribution per block for an example error-propagation frame selected from the 10000 simulated \myBlue{frames.\footnote{The example frames demonstrate the typical behavior of all the recorded error frames of a given type.}} Here, we see that the error propagation starts at block 606, which has 354 errors, and continues to the end of the frame. The number of bit errors in block 607 increases to around 1200. Then, in the remaining blocks, the number of bit errors is around 1500.

\begin{figure}[htbp]
\centering
\begin{minipage}[t]{0.485\textwidth}
\centering
\includegraphics[width=8cm]{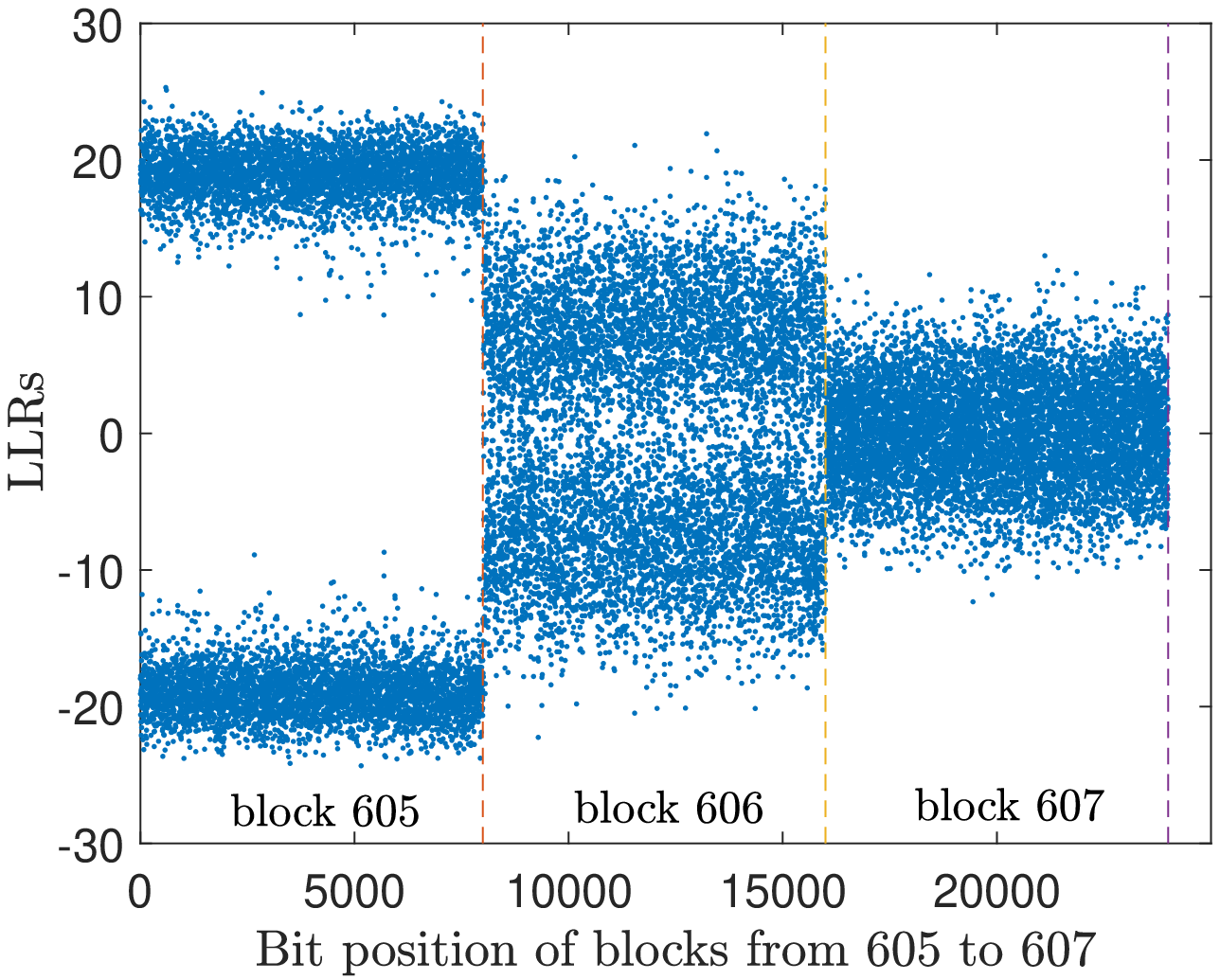}
\setlength{\abovecaptionskip}{-0.3cm} 
\caption{The LLRs of blocks 605, 606, and 607 in an example error-propagation frame from a rate $R=1/3$ SBCC with $T=8000$.}
\label{fig:LLRDist4917}
\end{minipage}
\hspace{0.1cm}
\begin{minipage}[t]{0.485\textwidth}
\centering
\includegraphics[width=8cm]{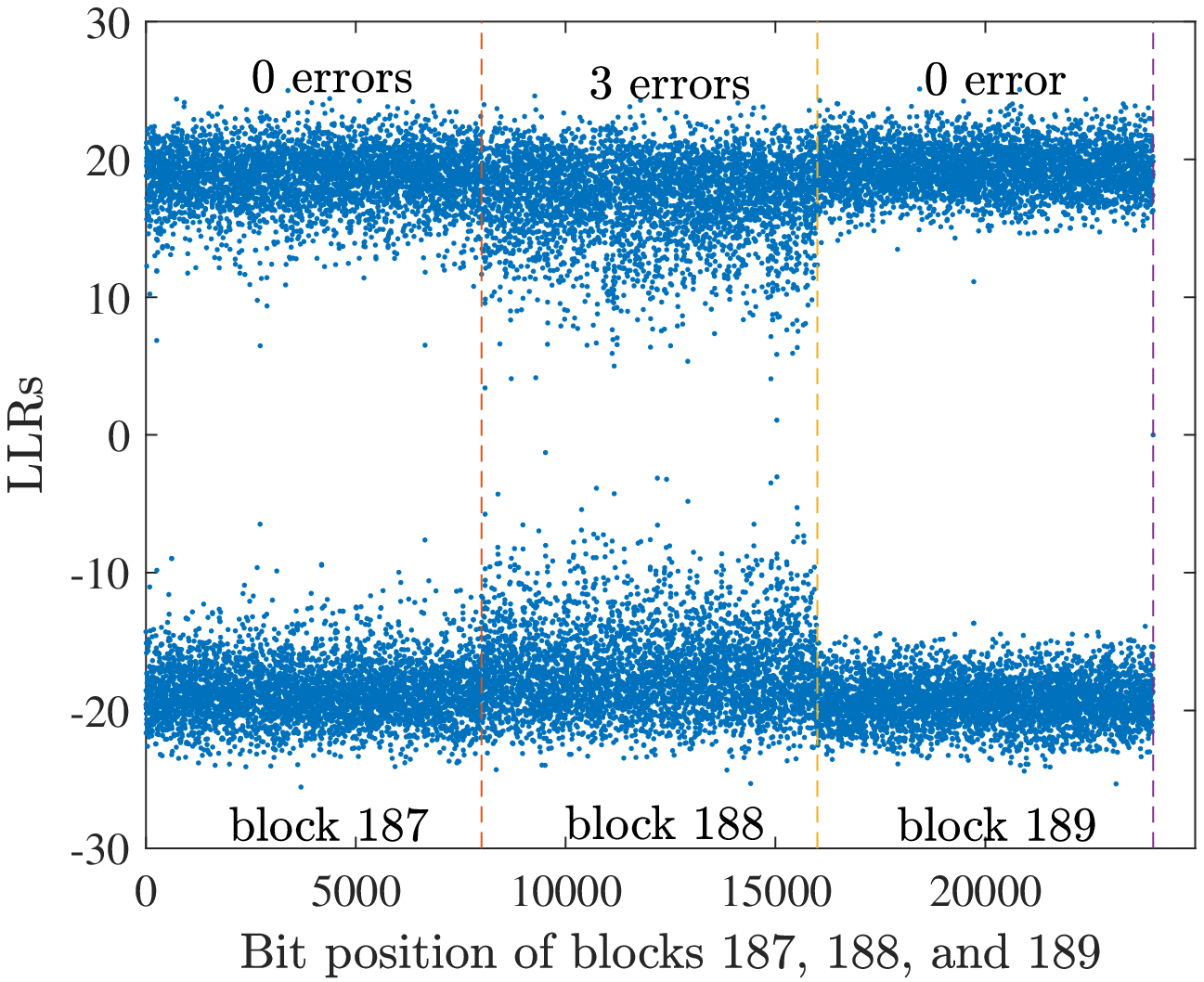}
\setlength{\abovecaptionskip}{-0.3cm} 
\caption{The LLRs of blocks 187, 188, and 189 in an example erroneous frame that does not display error propagation from a rate $R=1/3$ SBCC with $T=8000$.}
\label{fig:LLRDist39}
\end{minipage}
\vspace{-4mm}
\end{figure}

In \myBlue{Fig. \ref{fig:LLRDist4917}}, we show the decoded LLRs of blocks 605 (0 errors), 606 (354 errors), and 607 (1224 errors) of an example burst-error frame. We see that the LLR magnitudes of block 605 are mostly around 20, while the LLRs of block 606 range from about -10 to +10 almost uniformly, and the LLR magnitudes of block 607 are mostly around zero. This indicates that when error propagation begins, the average LLR magnitudes in a block quickly deteriorate to around zero, resulting in a sequence of unreliable blocks.


We now examine the bit error distribution per block in a typical erroneous frame that does not exhibit error propagation, selected from the same 10000 simulated frames. The example frame selected contains a total of 3 error bits confined to block 188. \myBlue{Fig. \ref{fig:LLRDist39}} shows the decoded LLRs of blocks 187 (0 errors), 188 (3 errors), and 189 (0 errors).
In this case, we see that a small number of bit errors in a single block does not trigger error propagation. In this regard, it is instructive to contrast the LLRs of block 606 in \myBlue{Fig. \ref{fig:LLRDist4917}}, which triggers error propagation, with those of block 188 in \myBlue{Fig. \ref{fig:LLRDist39}}, which does not. \hfill $\blacksquare$

In summary, for large block size $T$, a small number of bit errors in a block tends to affect only one or (occasionally) two blocks at a time, while larger numbers of bit errors in a block typically trigger error propagation. Also, when error propagation occurs, the corresponding decoded LLR magnitudes are highly unreliable, which indicates that we can design mitigation measures to detect and combat error propagation based on the decoded LLR magnitudes.

\begin{figure}
        \center
        \includegraphics[width=  \textwidth]{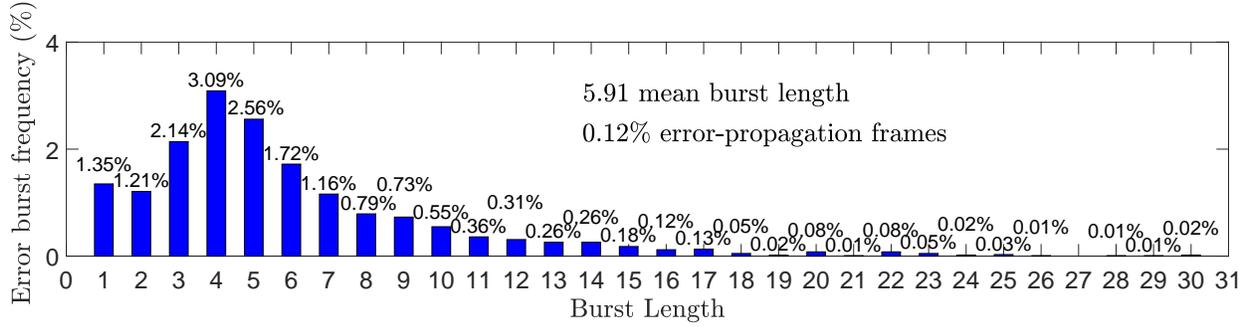}\\
        \vspace{-5mm}
        \caption{The frequency of the error frames in a rate $R=1/3$ SBCC with $T=100$, $w=14$, $I_1 = 1$, $I_2 = 20$, and $L=1000$ at $E_b/N_0 = 1.2~dB$.}
        \label{fig:ErrDistP100}
        \vspace{-6mm}
\end{figure}

\emph{\myBlue{Example 4}}: We next consider the case of a smaller permutor (block) size $T$. 10000 frames of the rate $R=1/3$ blockwise SBCC from Example 1 were simulated at $E_b/N_0 = 1.2$ dB (corresponding to a BER of about $10^{-4}$) with $T=100$, $w=14$, $I_1 = 1$, $I_2 = 20$, and $L=1000$. The frequency of the burst-error frames and error-propagation frames among the 10000 simulated frames, \myBlue{along with the mean burst length,} is shown in \myBlue{Fig. \ref{fig:ErrDistP100}}.
We see that, compared to using a larger permutor (block) size (see \myBlue{Fig. \ref{fig:BurstDistWinP8000}}), burst-error frames are in the majority, \myBlue{the burst errors are longer on average,} and there are relatively few error-propagation frames.

\begin{figure}
         \centering
            \includegraphics[width= \textwidth]{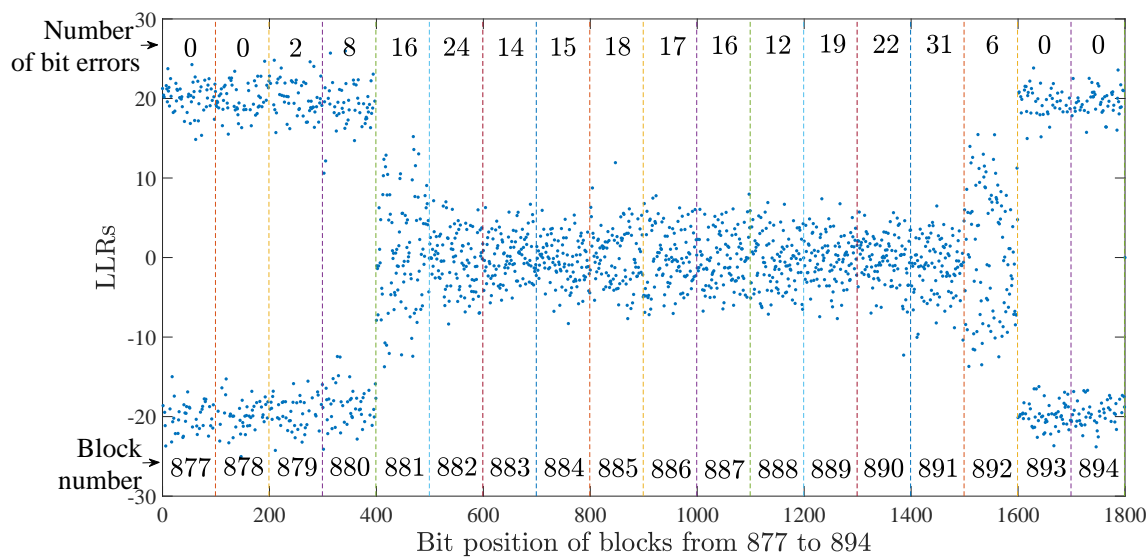}
  \setlength{\abovecaptionskip}{-1cm} 
  \caption{The bit error distribution per block and the LLRs of blocks 877 to 894 in frame 843 of a rate $R=1/3$ SBCC with $T=100$. }\label{fig:P100W14EbN12Fr843}
  \vspace{-5mm}
\end{figure}

We now examine a typical burst-error frame, which has burst length 14 (from block 879 to block 892), in more detail. \myBlue{Fig. \ref{fig:P100W14EbN12Fr843}} shows the bit error distribution per block along with the decoded LLRs. We see that in this case (with $T=100$), unlike the case in \myBlue{Fig. \ref{fig:ErrDist4917}} (with $T=8000$), the decoder recovers from the burst of block errors, and error propagation does not occur. However, the average magnitudes of the LLRs in the burst are relatively small (roughly between -10 to +10), which is similar to the LLR behavior shown in \myBlue{Fig. \ref{fig:LLRDist4917}} when $T=8000$.

\begin{figure}
    \centering
    \setlength{\subfigcapskip}{-0.5cm}
    \subfigure[Encoder 1]{
        \label{scatterplot:subfig:a}
        \includegraphics[width=0.45 \textwidth]{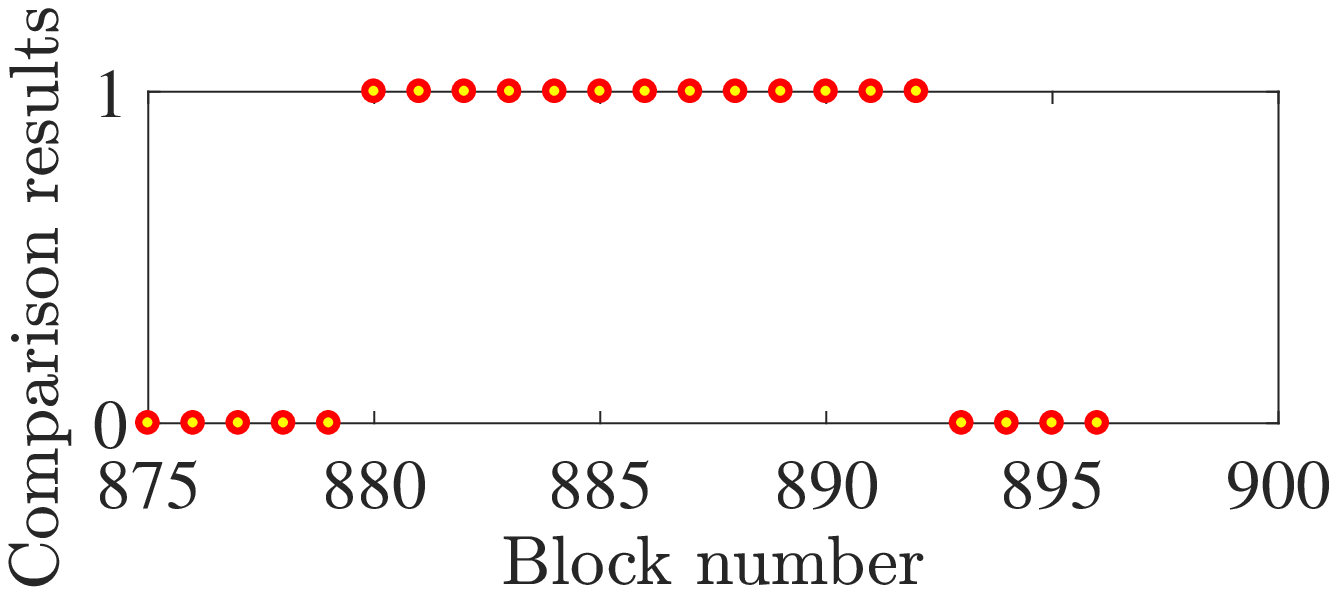}}
    \hspace{1cm}
    \subfigure[Encoder 2]{
            \label{scatterplot:subfig:b}
            \includegraphics[width=0.45 \textwidth]{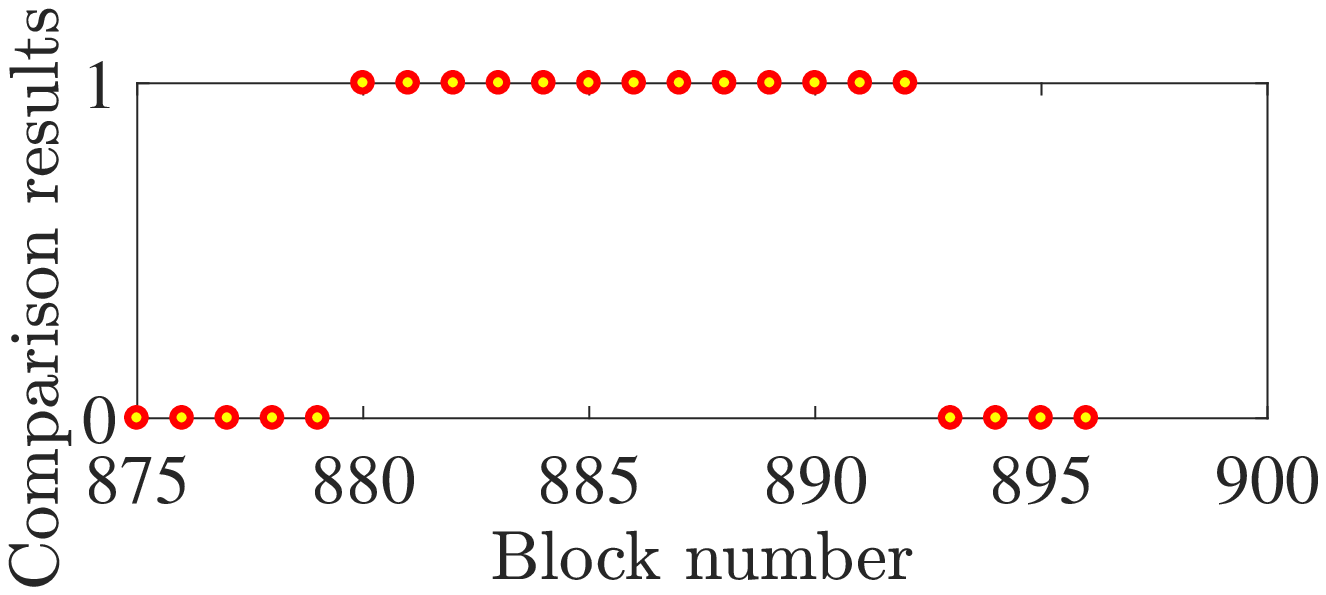}}
  \setlength{\abovecaptionskip}{-0.45cm} 
  \caption{The difference between the actual sequence of parity input blocks and the correct sequence of parity input blocks in each block (``1" represents ``different" and ``0" represents ``the same"). }\label{fig:DiffPrityIn}
  \end{figure}

\begin{figure}
    \centering
    \setlength{\subfigcapskip}{-0.5cm}
    \subfigure[Encoder 1]{
        \label{scatterplot:subfig:a}
        \includegraphics[width=0.45 \textwidth]{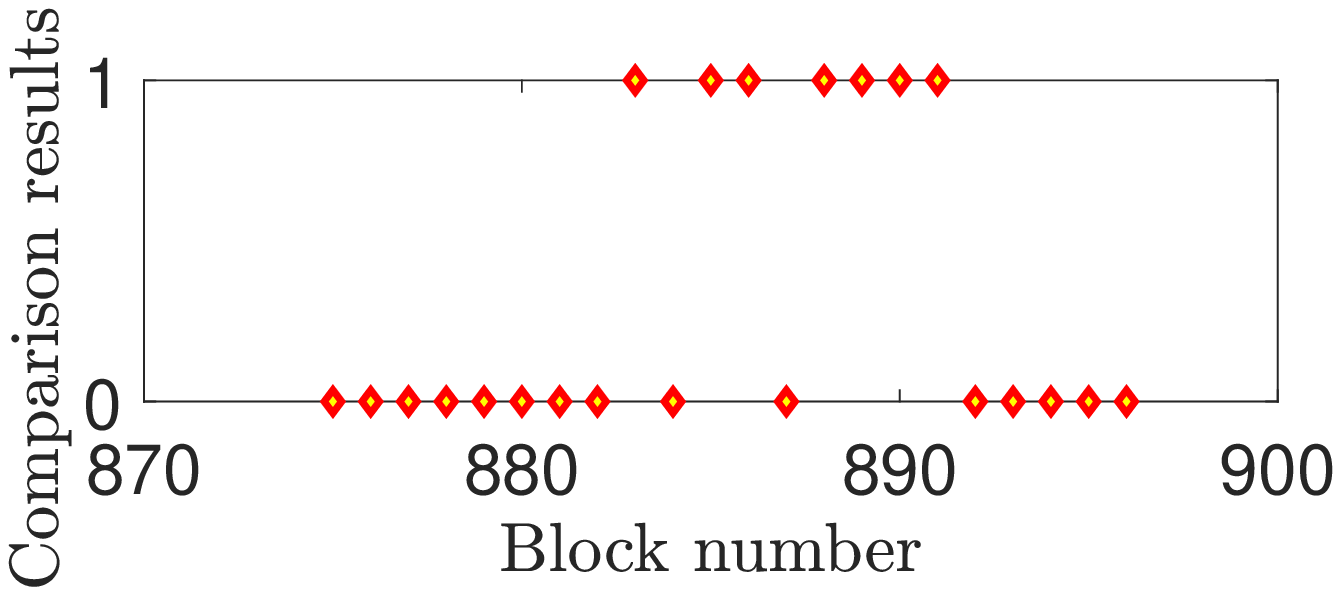}}
    \hspace{1cm}
    \subfigure[Encoder 2]{
            \label{scatterplot:subfig:b}
            \includegraphics[width=0.45 \textwidth]{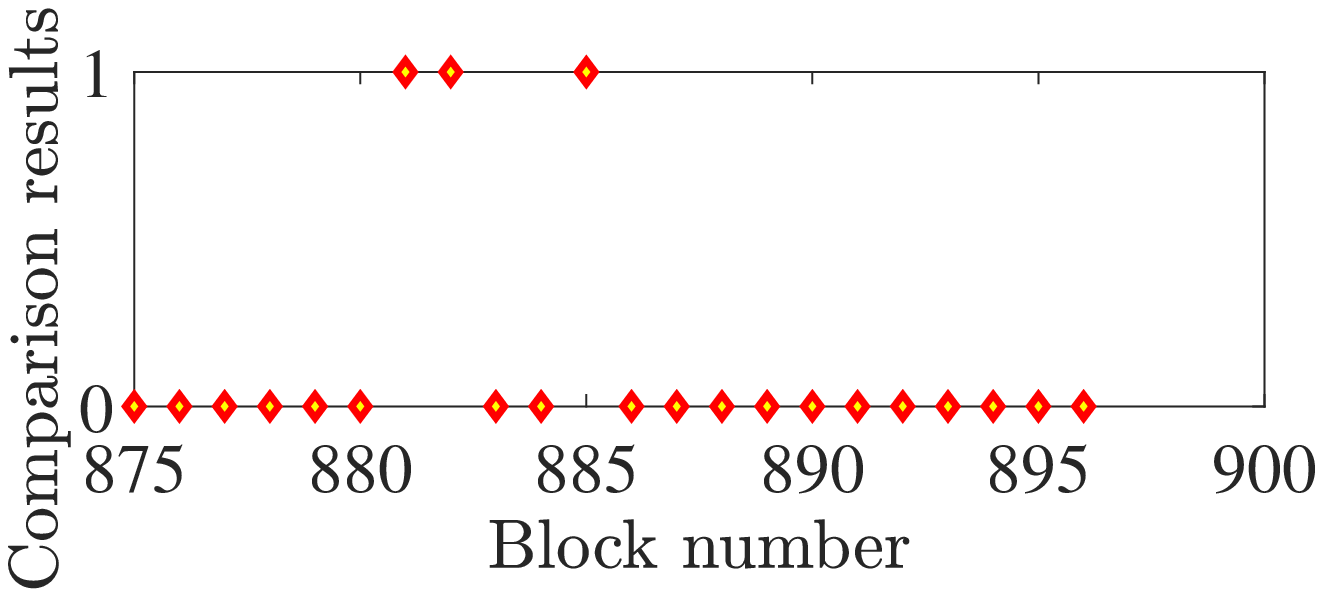}}
  \setlength{\abovecaptionskip}{-0.45cm} 
  \caption{The difference between the actual initial encoder state sequences and the correct initial encoder state sequences (``1" represents ``different" and ``0" represents ``the same"). }\label{fig:DiffState}
  \end{figure}

In order to better understand the process of decoder recovery from an error burst, we tracked the superstates obtained from the decoded sequence by making hard decisions on the LLRs of the two parity output blocks (parity inputs for the next block) and the two encoder states after each information block is decoded and compared them to the superstates obtained by encoding the correct sequence of information blocks.
\myBlue{Figs. \ref{fig:DiffPrityIn}-\ref{fig:DiffState}} show the comparative results of these two superstate sequences, where, in order to highlight the details of the burst-error blocks, we only show the results in their vicinity. (The superstates corresponding to the blocks not shown in \myBlue{Figs. \ref{fig:DiffPrityIn}-\ref{fig:DiffState}} are the same in both cases.) From \myBlue{Fig. \ref{fig:DiffPrityIn}}, we see that the parity input block portion of the superstate sequences differs from block 880 to block 892, which agrees exactly with the distribution of burst-error blocks. In other words, starting with block 879 and continuing through block 891, the hard decisions obtained from the parity output block LLRs of both component decoders are incorrect, causing incorrect parity input blocks in the succeeding blocks. \myBlue{Fig. \ref{fig:DiffState}} compares the initial encoder state portion of the superstate (obtained by making hard decisions on the final encoder state LLRs of the previous block) in the two cases. Here, the results are somewhat different, with encoder 1 having only 7 different initial states (out of the 13 error blocks), while encoder 2 has only 3 different initial states. In other words, the 100-bit initial parity input block portion of the superstate has a greater influence on the propagation of block errors than does the 2-bit initial encoder state portion, and error propagation only ends
when both the parity input blocks and the initial encoder states remerge. Also, although we see here that (particularly for small block sizes) bursts of block errors don't necessarily result in error propagation and the decoder can recover, additional burst-error blocks can occur later in a long frame or in a streaming application. \hfill $\blacksquare$

\myBlue{Examples 3 and 4} show that, for larger permutor (block) sizes, error propagation or single block error frames are the most likely, while for smaller permutor (block) sizes, burst-error frames occur more often. Therefore it is necessary to design mitigation techniques to combat both error propagation and burst errors. Based on the information obtained in \myBlue{Examples 3 and 4,} i.e., that the absolute values of the LLRs of the information bits decrease during error propagation or burst errors, algorithms can be designed to combat these error conditions.
In addition, it is important to be able to detect error propagation or a burst error early in the process, to avoid having to accept large numbers of decoded block errors. Therefore, the span of blocks over which the LLRs are observed must be carefully chosen.

In the following section, we present three techniques designed to mitigate error propagation and burst errors of finite duration.

\section{Error Propagation Mitigation}
In this section, we propose a window extension algorithm, a resynchronization mechanism, and a retransmission strategy to mitigate the effect of error propagation in sliding window decoding of SBCCs.

\subsection{Window Extension Algorithm}
In \cite{Zhu2017TCOM}, window decoding of SBCCs is performed with a fixed window size $w$. Based on the results presented in Fig. \ref{fig:DiffWin}, we now introduce a variable window size concept for sliding window decoding, where the window size can change from an initial value $w=w_{\rm{init}}$ to a maximum of $w = w_{\rm{max}}>w_{\rm{init}}$.
Before describing the \emph{window extension algorithm}, we give some definitions.
Let
${\bm{\ell}}^{\left( i,j \right)} = \left( {{\ell}_{0}^{\left( i,j \right)},{\ell}_{1}^{\left( i,j \right)},{\ell}_{2}^{\left( i,j \right)}, \ldots ,{\ell}_{T - 1}^{\left( i,j \right)}} \right)$ denote the decision LLRs of the $T$ information bits in the $i$th block, $i \in \left\{ {t,t + 1, \ldots ,t + w - 1} \right\}$, of the current window after the $j$th horizontal iteration. Then the \emph{average absolute LLR} of the $T$ information bits in block $i$ after the $j$th horizontal iteration is given by
\begin{equation}\label{eq:ld}
\bar {\ell}^{\left( {i,j} \right)} = \frac{1}{T}\sum\limits_{k = 0}^{T - 1} {\left| {{\ell}_{k}^{\left( i,j \right)}} \right|}.
\end{equation}
Also, we define the \emph{observation span} $\tau$ as the number of consecutive blocks in the decoding window over which the average absolute LLRs are to be examined.

During the decoding process, the window extension algorithm operates as follows: with $w=w_{\rm{init}}$, when the number of horizontal iterations reaches its maximum value $I_2$, if any of the average absolute LLRs of the first $\tau$ blocks in the current window, $1 \le \tau  \le w$, is lower than a predefined \emph{threshold} $\theta$, i.e., if
\begin{equation}\label{eq:2}
\bar {\ell}^{\left( {i,I_2} \right)} < {\theta}, ~~~~~\mathrm{for~any~i} \in \left\{ {t, t+1, \ldots, t+\tau-1} \right\},
\end{equation}
then the target block is not decoded, the window size is increased by 1, and the decoding process restarts with horizontal iteration number 1.\footnote{\myBlue{When decoding restarts, all the LLRs in the old blocks, except for the channel LLRs, are initialized to be 0s. In other words, the previous intermediate messages are not reused.}}
This process continues until either the target block is decoded or the window size reaches $w={w_{\max }}$, in which case the target block is decoded regardless of whether \myBlue{\eqref{eq:2}} is satisfied.

\begin{figure*}[ht]
        \center
    \includegraphics[width= \textwidth]{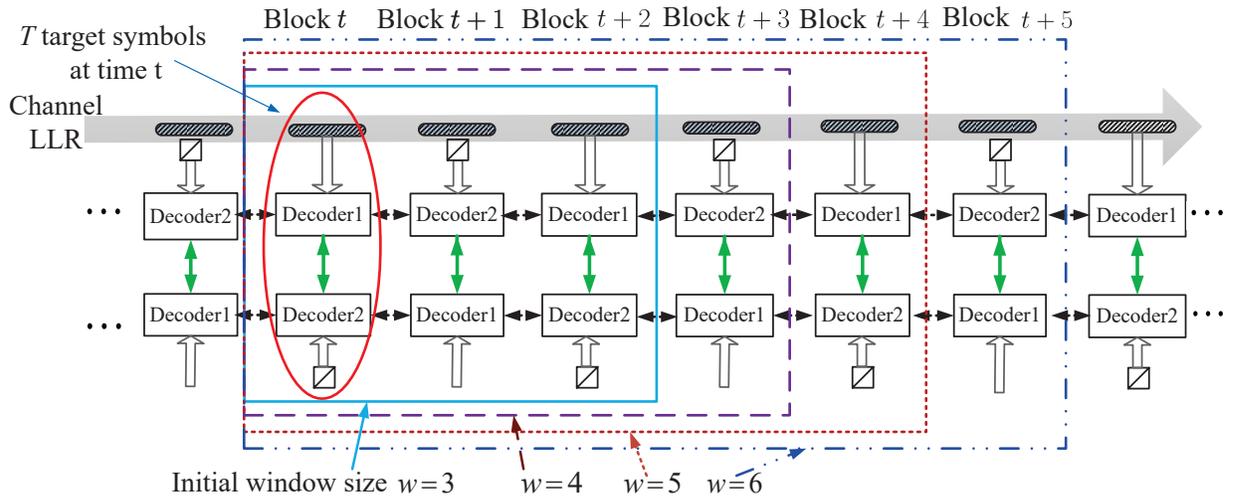}\\
    \setlength{\abovecaptionskip}{-0.1cm} 
    \caption{Sliding window decoder with the window extension algorithm.}
    \label{fig:WinExt}
    \vspace{-3mm}
\end{figure*}

Assuming an initial window size $w=w_{\rm{init}}=3$, \myBlue{Fig. \ref{fig:WinExt}} illustrates how the decoder window size increases by 1 each time \myBlue{\eqref{eq:2}} is satisfied, up to a maximum window size of \myBlue{$w=w_{\max}=6$.
Note that when window extension is triggered, the decoding delay, along with the decoding complexity, increases, so that an average latency measure must be adopted to characterize delay. Also, some buffering is required, and the decoder output is no longer continuous. These practical considerations suggest that $w_{\max}$ should not be too large.\footnote{\myBlue{Since, during horizontal iterations, messages from a given block are only shared with one adjacent block, the processing can be achieved, in principle, by using the existing hardware with a fixed window size $w=w_{\rm{init}}$ serially, along with additional memory, to increase ${w_{\max }}$ as needed}.} If error propagation persists given this constraint, window extension can be combined with one of the other mitigation methods, as discussed later in this section.}
Full details of the window extension algorithm are given in Algorithm \ref{Alg:WinExt_Alg} in the appendix.

\begin{figure}
    \centering
    \setlength{\subfigcapskip}{-0.6cm}
    \subfigure[$T=8000$, $w_{\rm{init}} = 3$, $w_{\max} = 6$, $\tau = 2$, and $\theta = 10$.]{
        \label{fig:BER_WinExt}
        \includegraphics[width= 0.47 \textwidth]{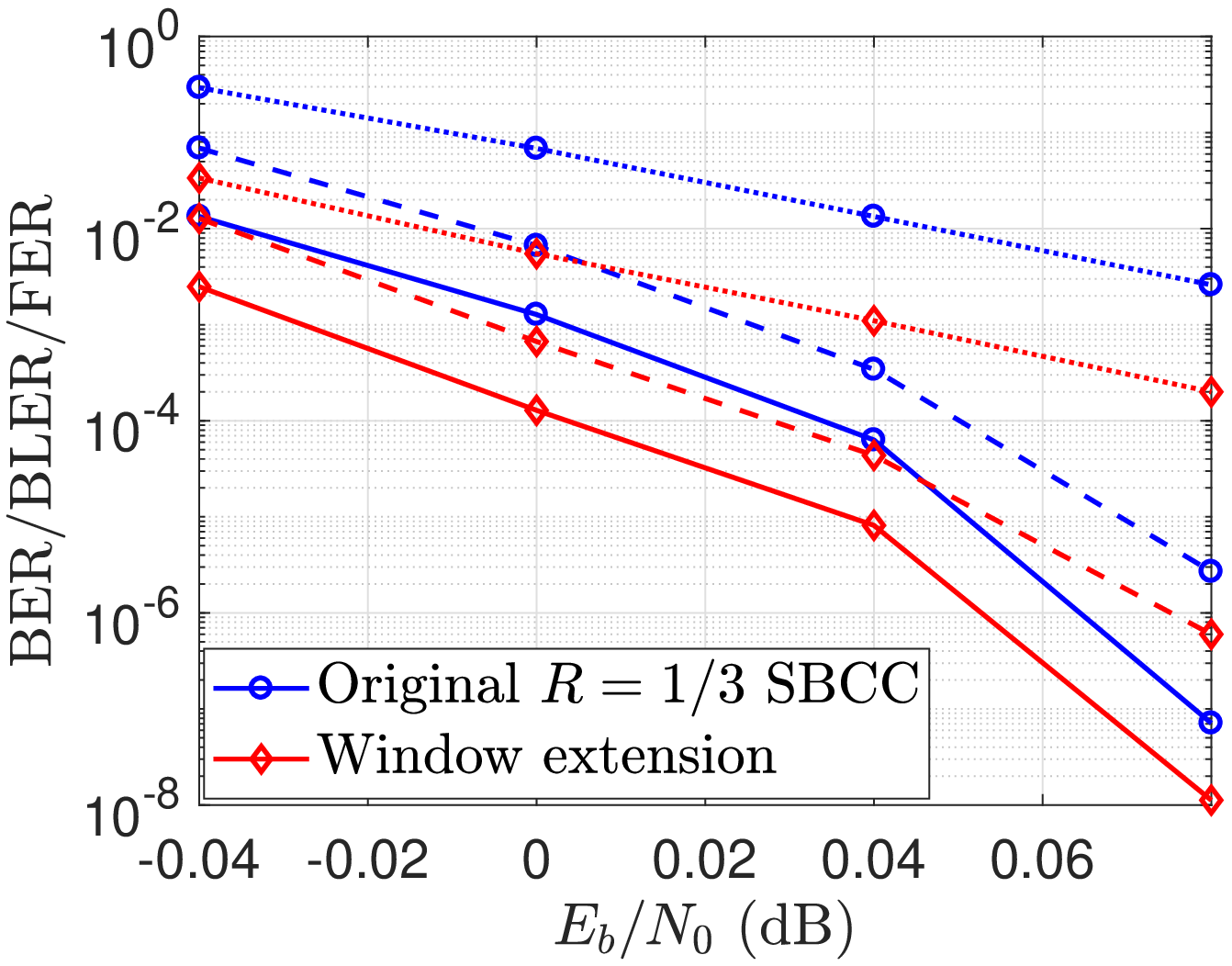}}
    \hspace{0.5mm}
    \subfigure[$T=500$, $w_{\rm{init}} = 6$, $w_{\max} = 12$, $\tau = 3$, $\theta = 10$.]{
            \label{fig:BER_WinExtT500}
            \includegraphics[width= 0.47 \textwidth]{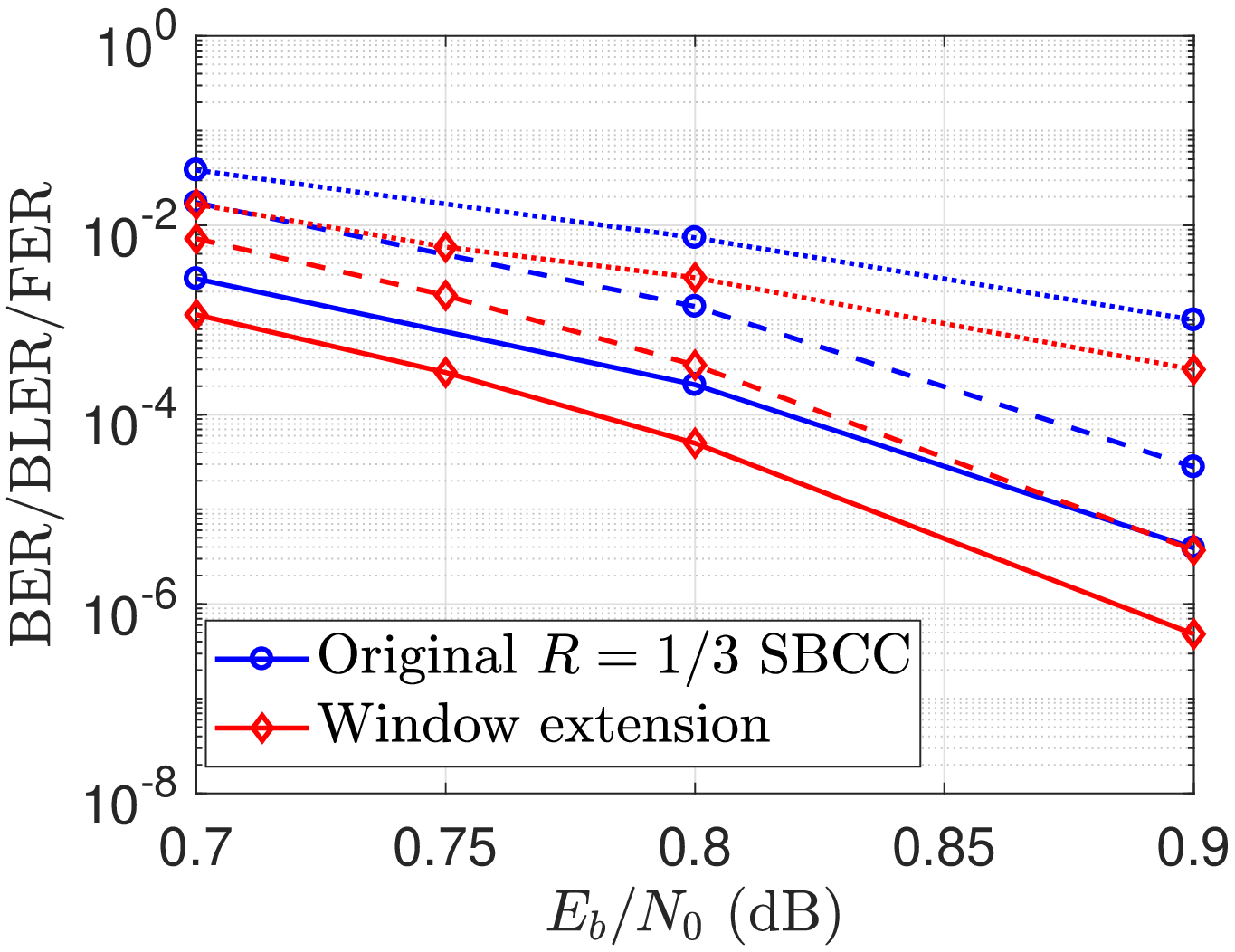}}

  \caption{BER (solid curves), BLER (dashed curves), and FER (dotted curves) performance comparison of a rate $R=1/3$ SBCC with and without the window extension.}\label{fig:TaoWinP8000}
  \vspace{-2mm}
  \end{figure}

For the same simulation parameters used in Example 1, the BER, BLER, and FER performance of a rate $R=1/3$ blockwise SBCC both with and without the window extension algorithm is shown in \myBlue{Fig. \ref{fig:BER_WinExt}}, where $T = 8000$, $w_{\rm{init}} = 3$, $w_{\max} = 6$, the observation span $\tau = 2$, and the threshold \myBlue{$\theta = 10$.\footnote{\myBlue{After some experimentation, $w_{\max}=6$ was found to give a reasonable tradeoff among complexity, memory requirements, and delay in this example.}}\footnote{\myBlue{The choice of $\theta = 10$ is based on the information regarding typical LLR magnitudes during error bursts and error propagation presented in Figs. \ref{fig:LLRDist4917} and \ref{fig:P100W14EbN12Fr843}. Note that the higher the threshold $\theta$, the more often window extension is triggered, which increases decoding complexity, while smaller values of $\theta$ risk failing to detect error propagation.}}} (Throughout the remainder of this section, we assume $I_1 = 1$, $I_2 = 20$, and $L = 1000$.) We see that window extension shows an order of magnitude improvement in BER, BLER, and FER compared to using a fixed window size. We also remark that, even though $w_{\max} = 6$, the average window size $\bar w$ is found to be only slightly larger than $w_{\rm{init}}$, e.g., $\bar w = 3.0014$ for ${E_b}/{N_0} = 0.04$ dB, since window extension is only activated in the few cases when error propagation is detected.

To examine the effect of a smaller block size, the BER, BLER, and FER performance of the rate $R=1/3$ blockwise SBCC of Example 1 both with and without the window extension is shown in \myBlue{Fig. \ref{fig:BER_WinExtT500}} for $T = 500$, $w_{\rm{init}} = 6$, $w_{\max} = 12$, $\tau = 3$, and $\theta = 10$. We again see that window extension shows almost an order of magnitude improvement in BER, BLER, and FER compared to using a fixed window size, and the average window size, e.g.,  $\bar w = 6.00004$ for ${E_b}/{N_0} = 0.9$ dB, is only slightly larger than $w_{\rm{init}}=6$.

To further illustrate the performance gains achieved by the window extension algorithm, the frequency of the burst-error frames and error-propagation frames over a total of 10000 frames, \myBlue{along with the mean burst length,} is shown in \myBlue{Fig. \ref{fig:TaoWinP8000}} for $E_b/N_0 = 0.04$ dB and $T=8000$. In this case, compared to \myBlue{Fig. \ref{fig:BurstDistWinP8000}}, we see that window extension reduces the frequency of both error propagation frames and length 1 \myBlue{burst-error} frames by roughly a factor of 10, while completely eliminating the small number of bursts of length 2. Also, we have observed empirically that the frequency of error frames decreases as we increase the observation span $\tau$. Therefore, in order to maintain an acceptable tradeoff between performance and decoding complexity,\myBlue{\footnote{Increasing $\tau$ also increases the complexity of performing the threshold test in \myBlue{(\ref{eq:2})}.}} we typically choose
\begin{equation}
\tau {\rm{ = }}\left\lceil {\frac{{{w_{{\rm{init}}}}}}{{\rm{2}}}} \right\rceil.
\label{eq:3}
\end{equation}


%

  \begin{figure}
    \centering

    \includegraphics[width= 0.6 \textwidth]{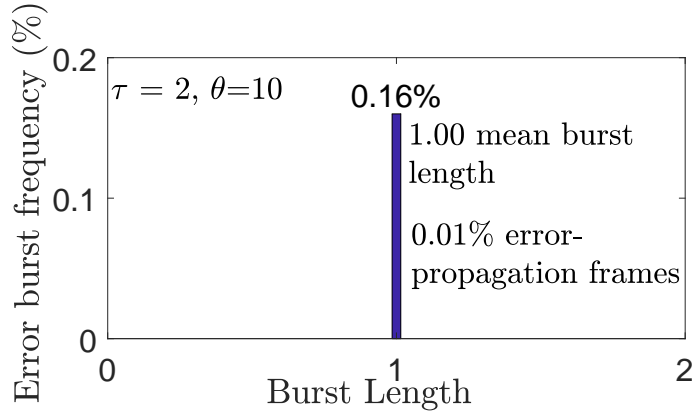}
  \setlength{\abovecaptionskip}{-0.15cm} 
  \caption{The frequency of the error frames in a rate $R=1/3$ SBCC with window extension for $T=8000$, $w_{\rm{init}}=3$, and $w_{\max}=6$ at $E_b/N_0 = 0.04$ dB.}
  \label{fig:TaoWinP8000}
  \vspace{-8mm}
  \end{figure}

\begin{figure}
    \centering
    \setlength{\subfigcapskip}{-0.6cm}
    \subfigure[Window decoding of an SBCC without window extension, $T = 500$, $w= 6$]{
        \label{scatterplot:subfig:a}
        \includegraphics[width=  \textwidth]{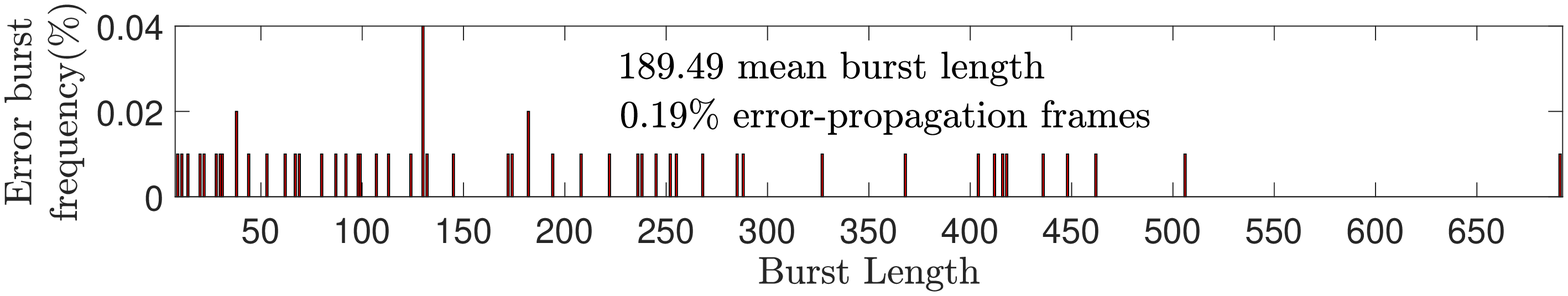}
        }
    \hfill
    \hfill
    \subfigure[Window decoding of an SBCC with window extension, $T = 500$, $w_{\rm{init}} = 6$, $w_{\max} = 12$, $\tau = 2$, $\theta = 10$]{
            \label{scatterplot:subfig:b}
            \includegraphics[width= \textwidth]{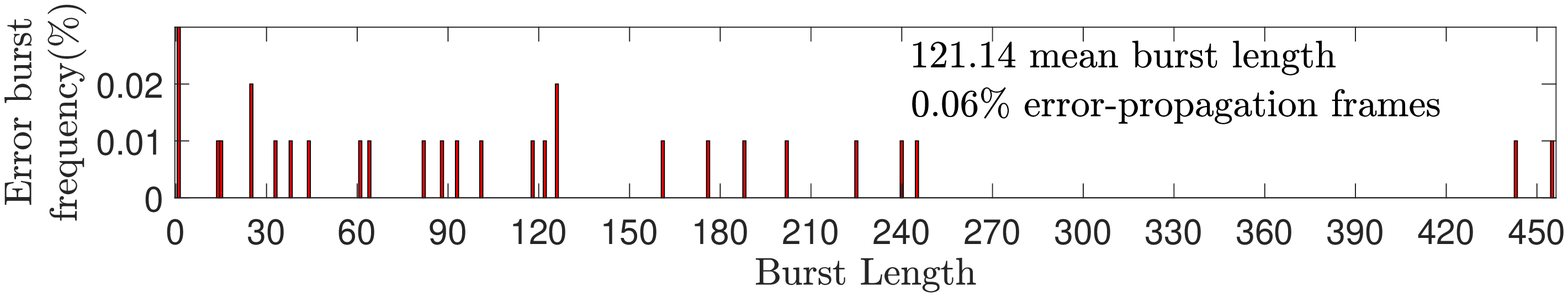}}
    \hfill
    \hfill
    \subfigure[Window decoding of an SBCC with window extension, $T = 500$, $w_{\rm{init}} = 6$, $w_{\max} = 12$, $\tau = 3$, $\theta = 10$]{
            \label{scatterplot:subfig:c}
            \includegraphics[width= \textwidth]{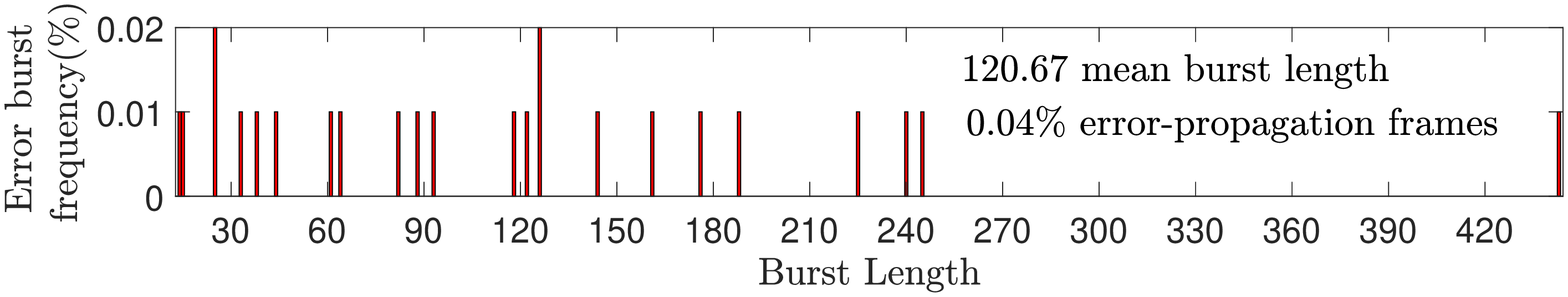}}
  \setlength{\abovecaptionskip}{-0.4cm} 
  \caption{The frequency of the error frames in a rate $R=1/3$ SBCC with and without window extension for $T=500$ at $E_b/N_0=0.8$ dB.}\label{fig:TaoWinP500}
  \vspace{-6mm}
  \end{figure}

To again examine the effect of a smaller block size, \myBlue{Fig. \ref{fig:TaoWinP500}} shows the frequency of the burst-error frames and error-propagation frames over a total of 10000 frames, \myBlue{along with the mean burst length,} both with and without window extension, for $T=500$ and $E_b/N_0 = 0.8$ dB. Plots are included for two different values of the observation span $\tau$, $\tau=2$ and $\tau=3$. Unlike the large block size ($T=8000$) case, we see here that window decoding results in many different burst-error lengths. (More detailed information about the error frames is given in Table \ref{TAB:ErrDistT500w6}, where any frame containing error propagation is counted as an error-propagation frame and the number of burst-error frames includes those with both single and multiple burst errors.) In particular, without window extension, we experience burst errors as long as 691 blocks, \myBlue{a mean burst of 189.49, and} 19 error-propagation frames. With window extension, the total number of burst-error frames, the maximum length of error bursts, \myBlue{the mean burst length,} and the number of error-propagation frames are all reduced, with $\tau=3$ performing better than $\tau=2$, consistent with our choice in \myBlue{(\ref{eq:3})}.

\begin{table}[ht]
\large
 \newcommand{\tabincell}[2]{\begin{tabular}{@{}#1@{}}#2\end{tabular}}
 \vspace*{-1mm}
 \caption{The distribution of error frames for a rate $R=1/3$ SBCC with and without window extension for $T = 500$ and $E_b/N_0=0.8$ \upshape dB.}
 \label{TAB:ErrDistT500w6}
 \vspace*{-15mm}
 \begin{center}
 \scalebox{0.8}{
\begin{tabular}{|c|c|c|c|c|c|}
\hline
\hline
 $ $ & \tabincell{c}{Number~of~\\error~frames}  & \tabincell{c}{Number~of~error-\\propagation~frames}  & \tabincell{c}{Number~of burst-\\error~frames}  & \tabincell{c}{Largest \\burst size} & \tabincell{c}{\myBlue{Mean} \\\myBlue{burst size}}  \\ \hline
 No~window~extension  & 74 & 19  & 55  & 691 & \myBlue{189.49}\\ \hline
 ${\tau} = 2$ & 35 & 6 & 29 & 455 & \myBlue{121.14} \\ \hline
 ${\tau} = 3$ & 28 & 4 & 24 & 443 & \myBlue{120.67} \\ \hline
 \hline
 \end{tabular}}
 \end{center}
 \vspace*{-8mm}
\end{table}

\begin{figure}
    \centering
   \includegraphics[width=  \textwidth]{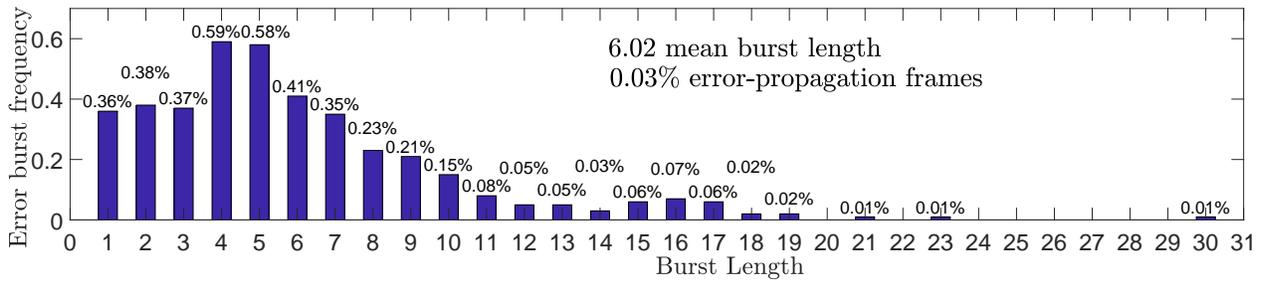}
   \setlength{\abovecaptionskip}{-0.7cm} 
  \caption{The frequency of the error frames in a rate $R=1/3$ SBCC with window extension for $T=100$ at $E_b/N_0 = 1.2$ dB.}
  \label{fig:TaoWinExtP100}
  \vspace{-2mm}
  \end{figure}

Considering the effect of an even smaller block size, \myBlue{Fig. \ref{fig:TaoWinExtP100}} shows the \myBlue{frequency} of the burst-error frames and error-propagation frames over a total of 10000 frames, \myBlue{along with the mean burst length,} with window extension for $T=100$ and $E_b/N_0 = 1.2$ dB  with $w_{\rm{init}}=14$ and $w_{\max}=20$. Comparing to \myBlue{Fig. \ref{fig:ErrDistP100}} without window extension, we see that window extension reduces the frequency of error-propagation frames from $0.12\%$ to $0.03\%$ \myBlue{and} the frequency of burst-error frames by about a factor of \myBlue{4, while the mean burst length stays about the same}.

\subsection{Resynchronization Mechanism}
\label{Sec:Resyn}
We see from \myBlue{Fig. \ref{fig:BER_WinExt}} that the window extension algorithm greatly reduces the effect of error propagation. However, for very long frames or for streaming applications, even one occurrence of error propagation can be catastrophic. We now introduce a resynchronization mechanism to address this problem.\myBlue{\footnote{Resynchronization can be employed with or without window extension. Resynchronization is considered without window extension in Section \ref{Sec:Resyn} and with window extension in Section \ref{Sec:WinRysn}.}}

As noted above, the parity input sequences in the first block of an SBCC encoder output sequence are known. Therefore, the input LLRs for the first block are more reliable than for the succeeding blocks. Motivated by this observation, and assuming the availability of an instantaneous noiseless binary feedback channel, we propose that, when the sliding window decoding algorithm is unable to recover from error propagation, the encoder resets to the $0$ state and restarts encoding. This resynchronization mechanism is described below.

In attempting to decode the target block at time $t$ in the window decoding algorithm, if the average absolute LLRs of the target block satisfy,
\begin{equation}\label{eq:Reset}
\setlength{\abovedisplayskip}{1pt}
\setlength{\belowdisplayskip}{1pt}
\bar {\ell}^{\left( t,I_2 \right)} < {\theta},
\end{equation}
we consider the target block as \emph{failed}, where $\theta$ is the same predefined threshold employed in window extension. If we experience $N_r$ consecutive failed target blocks, we then declare an error propagation condition and initiate encoder and decoder resynchronization using the feedback channel. In other words, the encoder 1) sets the initial states of the two component convolutional encoders to ``0'', and 2) begins encoding the next block with two known (all ``0'') parity input sequences together with the next information block. Meanwhile, the decoder makes decisions based on the current LLRs for the $w$ blocks in the current window and restarts decoding once $w$ new blocks are received.
Full details of the resynchronization mechanism are given in Algorithm \ref{Alg:Reset_Alg} in the appendix.

%

\begin{figure}
    \centering
    \setlength{\subfigcapskip}{-0.4cm}
    \subfigure[$T=8000$, $w=3$, and $N_r = 2$.]{
        \label{fig:P8000W3Rys}
        \includegraphics[width= 0.47 \textwidth]{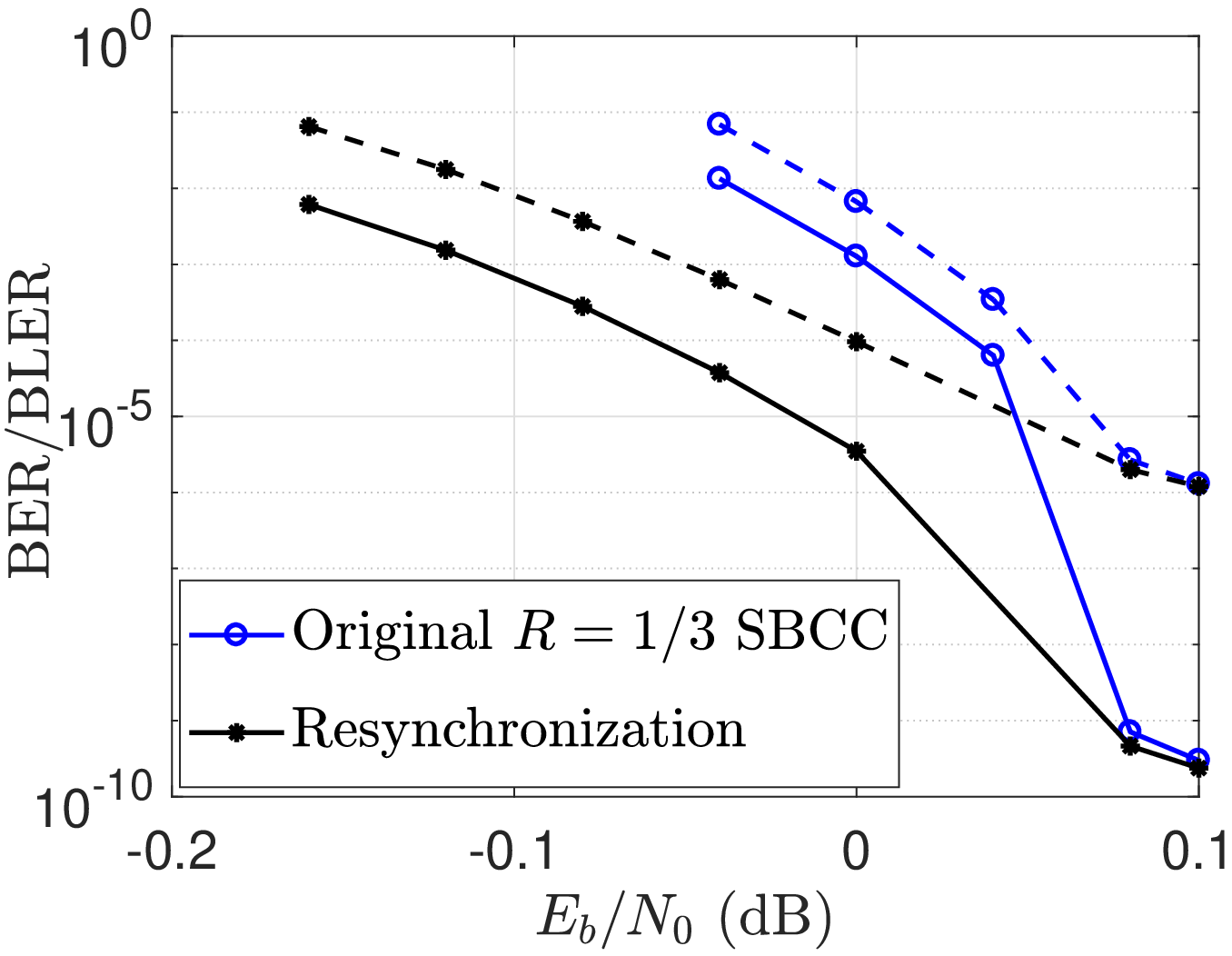}}
    \hspace{0.1cm}
    \subfigure[$T=500$ and $w=6$ for $N_r = 1$ and $N_r = 2$.]{
            \label{fig:P500W6RysNr}
            \includegraphics[width= 0.47 \textwidth]{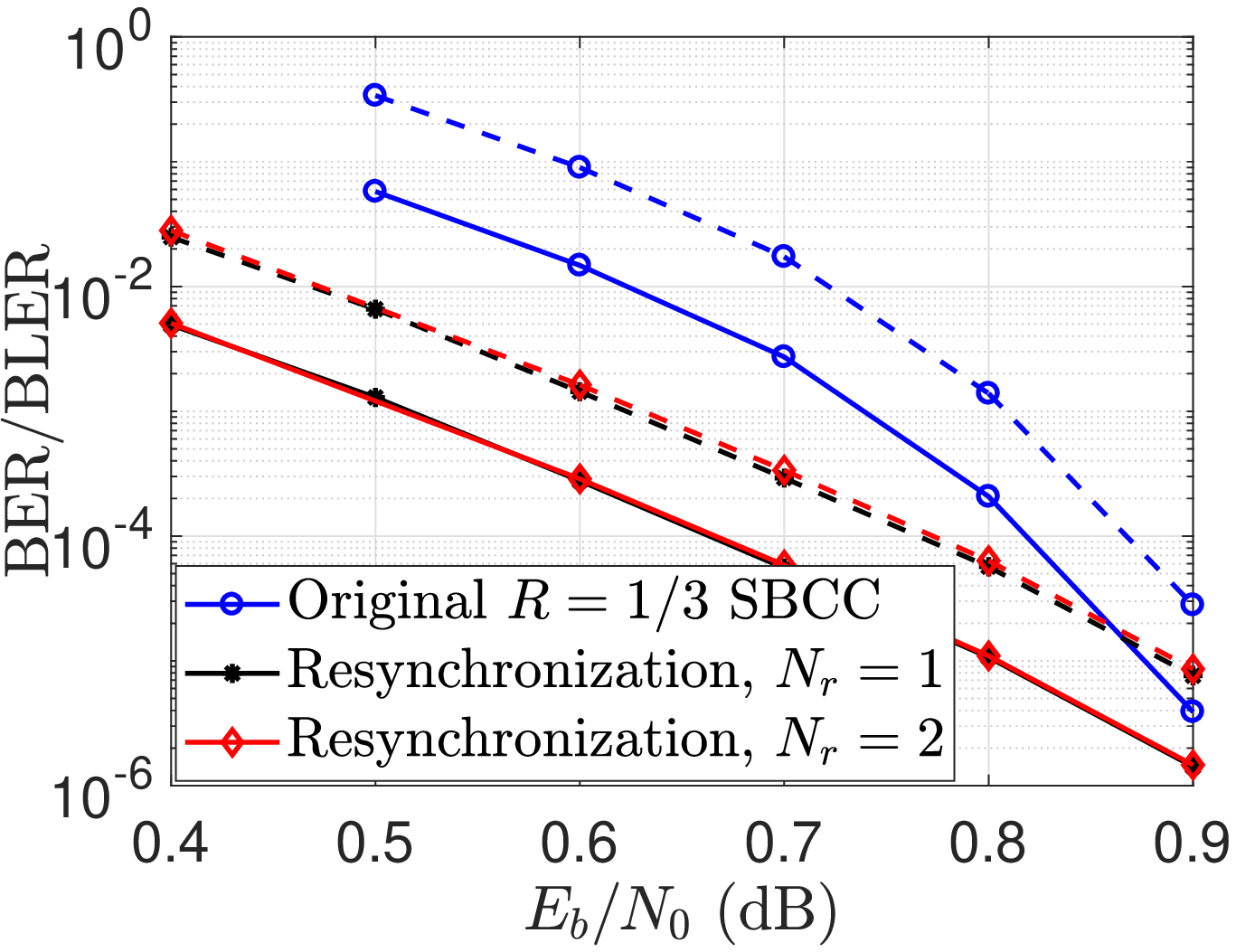}}
  \caption{BER (solid curves) and BLER (dashed curves) comparison of a rate $R=1/3$ SBCC with and without resynchronization.}\label{fig:P8000P500Rys}
  \vspace{-5mm}
  \end{figure}

In order to test the efficiency of resynchronization, we simulated the rate $R=1/3$ blockwise SBCC of Example 1 with different permutor (block) sizes and different numbers of consecutive failed target blocks ($N_r$). \myBlue{Fig. \ref{fig:P8000W3Rys}} shows the BER/BLER performance comparison with and without the resynchronization.\myBlue{\footnote{Although resynchronization terminates error propagation in a frame, thus improving both the BER and the BLER, it does not reduce the number of frames in error. For this reason, FER results are not included in \myBlue{Figs. \ref{fig:P8000W3Rys} and \ref{fig:P500W6RysNr}}.}} The parameters are $T=8000$, $w=3$, and $N_r = 2$. We see that, with the help of resynchronization, we obtain about two orders of magnitude improvement in both the BER and the BLER in the typical SNR operating range.\myBlue{\footnote{Note that \myBlue{Fig. \ref{fig:TaoWinP8000}} implies that $N_r=1$ would not be a good choice here, since the high frequency of single block errors would result in only modest improvements in BER/BLER at a cost of significantly more resynchronization requests, i.e., increased decoding complexity.}} We also note that the curves tend to merge as the SNR increases, since error propagation, and thus the need for window extension or resynchronization, is rare under good channel operating conditions.

\myBlue{Fig. \ref{fig:P500W6RysNr}} shows the BER/BLER performance comparison with resynchronization for two different values of $N_r$, with $T=500$ and $w=6$. We see that the performance with $N_r = 1$ is slightly better than with $N_r = 2$, which implies that, for short block lengths, resynchronization should be launched as soon as \myBlue{\eqref{eq:Reset}} is satisfied by a single target block.\myBlue{\footnote{\myBlue{Fig. \ref{scatterplot:subfig:a}} implies that $N_r = 1$ is a good choice here because of the relative scarcity of single block errors.}}
\subsection{Window Extension plus Resynchronization}
\label{Sec:WinRysn}
Window extension and resynchronization can also be employed together in order to further mitigate the effects of error propagation. Basically, window extension is triggered whenever \myBlue{\eqref{eq:2}} is satisfied. When the window size $w$ reaches $w_{\max}$ and  \myBlue{\eqref{eq:2}} is still satisfied, the decoder resets $w$ to $w_{\rm{init}}$ and then checks if \myBlue{\eqref{eq:Reset}} is satisfied. If so, resynchronization is launched.
Algorithm \ref{Alg:WinReset_Alg} in the appendix gives the details of window extension plus resynchronization.

\begin{figure}
    \centering
    \setlength{\subfigcapskip}{-0.4cm}
    \subfigure[]{
            \label{fig:BER_WinExtReset}
            \includegraphics[width= 0.47 \textwidth]{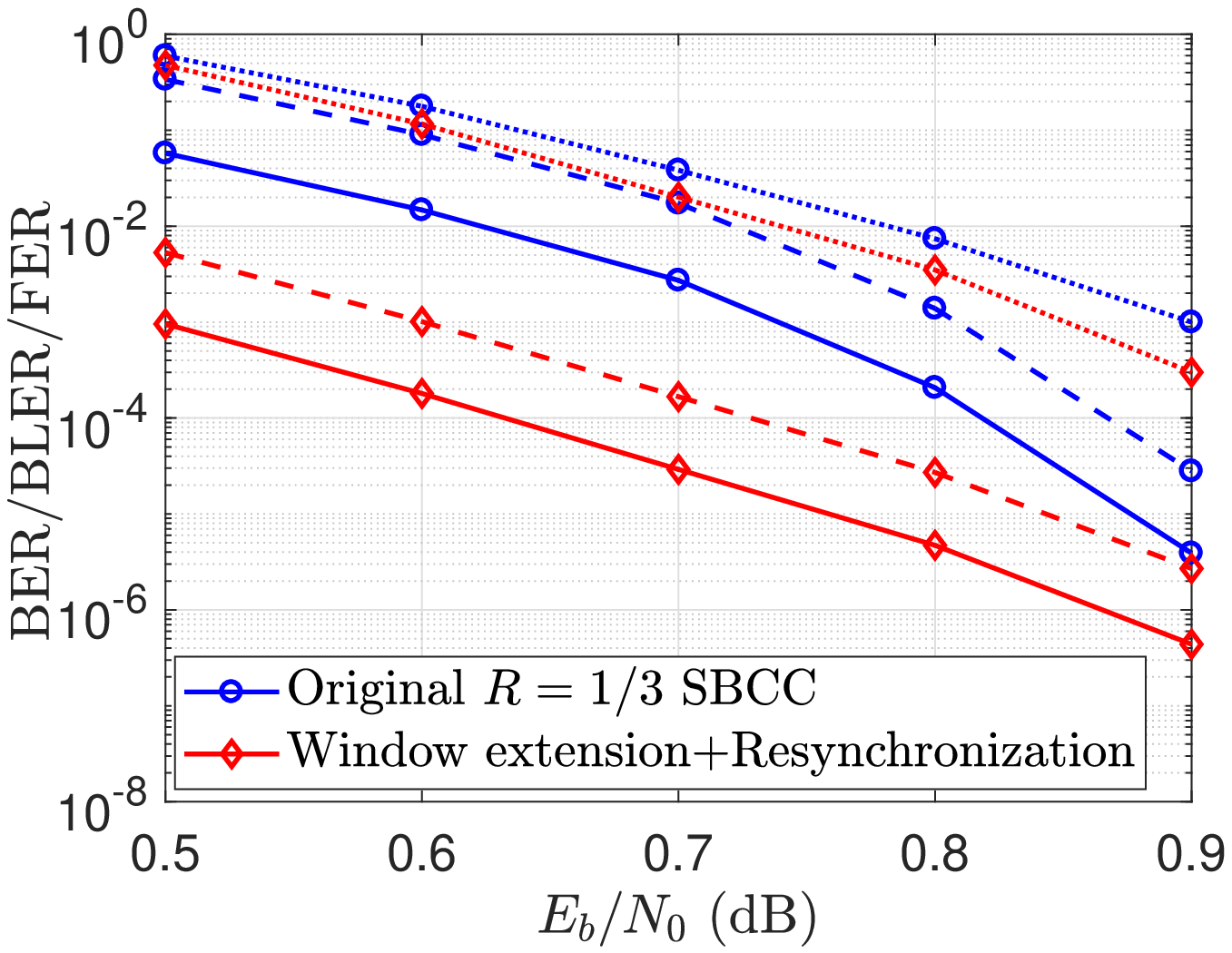}}
    \hspace{0.1cm}
    \subfigure[]{
        \label{fig:P500W6ExtReTran}
        \includegraphics[width= 0.47 \textwidth]{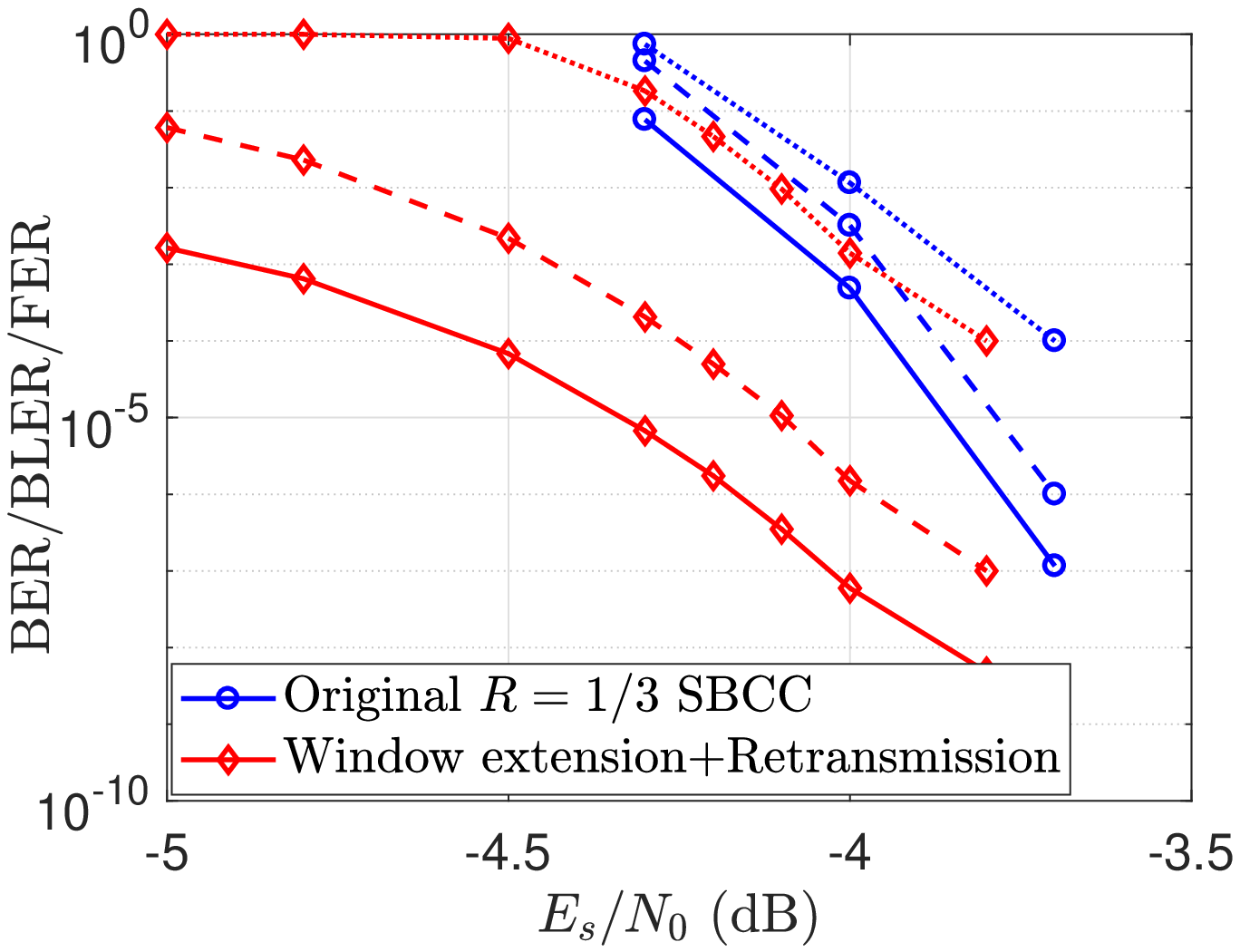}}
  \caption{BER (solid curves), BLER (dashed curves), and FER (dotted curves) comparison of a rate $R=1/3$ SBCC with window extension combined with (a) resynchronization and (b) retransmission.}\label{fig:WinReTran_Reset}
  \vspace{-7mm}
  \end{figure}

To demonstrate the efficiency of resynchronization combined with window extension, the BER, BLER, and FER performance of a rate $R=1/3$ blockwise SBCC employing both techniques is shown in \myBlue{Fig. \ref{fig:BER_WinExtReset}} for $T=500$, $N_r = 2$, $w_{\rm{init}}=6$, $w_{\max}=12$, $\tau = 2$, and $\theta = 10$. We see that, compared to the $R=1/3$ blockwise SBCC of Example 1, the rate $R=1/3$ blockwise SBCC with window extension and resynchronization gains approximately two orders of magnitude in BER and BLER and about one order of magnitude in FER at typical operating SNRs.\myBlue{\footnote{Including window extension along with resynchronization allows improvements in the FER, unlike the results for resynchronization alone.}} We also note that, comparing to \myBlue{Fig. \ref{fig:P500W6RysNr}}, combining resynchronization with window extension gains almost an order of magnitude in BER and BLER compared to resynchronization alone.

\subsection{Retransmission Strategy}
In the resynchronization mechanism, once resynchronization is triggered, decisions are made on the remaining blocks in the current window, where it is likely that errors still exist. In order to eliminate these errors, we now describe a retransmission strategy as an alternative to resynchronization.

After a target block is decoded, if its average absolute LLRs satisfy \myBlue{\eqref{eq:Reset}}, we consider the target block as \emph{failed}. If there are ${N'_r}$ consecutive failed target blocks, retransmission is triggered, again employing an instantaneous noiseless binary feedback channel, using the following steps:
\begin{itemize}
\item The encoder sets the initial states of the two component convolutional encoders to ``0'';
\item The information blocks corresponding to the ${N'_r}$ failed blocks and the $w-1$ remaining blocks in the window reenter the encoder, in sequence, and the corresponding encoded blocks are retransmitted.\myBlue{\footnote{This requires a buffer at the transmitter to store the most recent ${N'_r}+w-1$ encoded blocks, so they are available for re-encoding when a retransmission request is received.}} The first retransmitted information block is encoded with two known (all ``0'') parity input sequences;	
\item The decoder is reset to its original state and decoding begins again with the first retransmitted block.
\end{itemize}

The details of this procedure are given in Algorithm \ref{Alg:Retran_Alg} in the appendix.

The difference between resynchronization and retransmission is that no blocks are retransmitted in the former case, whereas ${N'_r}+w-1$ blocks are retransmitted at a time in the latter case. Therefore, unlike resynchronization, retransmission involves some rate loss. \myBlue{However, unlike a conventional hybrid automatic repeat request (HARQ) scheme, the parity feedback (memory) in the encoding process and the fact that the component encoder states are reset to zero results in a different sequence of transmitted blocks (albeit representing the same sequence of information blocks), meaning that techniques such as selective repeat and Chase combining cannot be employed.\footnote{\myBlue{We choose to reset the component encoders to the ``0'' state because BCCs are a type of spatially coupled code and thus benefit from termination at the beginning of a frame. It would also be possible to not reset and selectively repeat only blocks that satisfy \eqref{eq:Reset}, thus improving throughput at a cost of reduced performance. As suggested by a reviewer, this would be an interesting option to investigate in future research.}}}
The \emph{average effective rate} (or \emph{throughput}) of the retransmission strategy is given by
\begin{equation}
\tilde R = \frac{{T\cdot {L}}}{{T/R\cdot\left( {L + {\bar S_r} \cdot \left( {{N'_r} + w - 1} \right)} \right)}},
\end{equation}
where $R$ is the code rate of the SBCC without retransmission and ${\bar S_r}$ is the average number of retransmissions in a frame.

In the following, we give two examples to illustrate the effectiveness of the retransmission strategy.

\begin{figure}
    \centering
    \setlength{\subfigcapskip}{-0.3cm}
    \subfigure[$T=8000$, ${N'_r} = 2$, and $w=3$.]{
        \label{fig:BER_ResetRateloss}
        \includegraphics[width= 0.47 \textwidth]{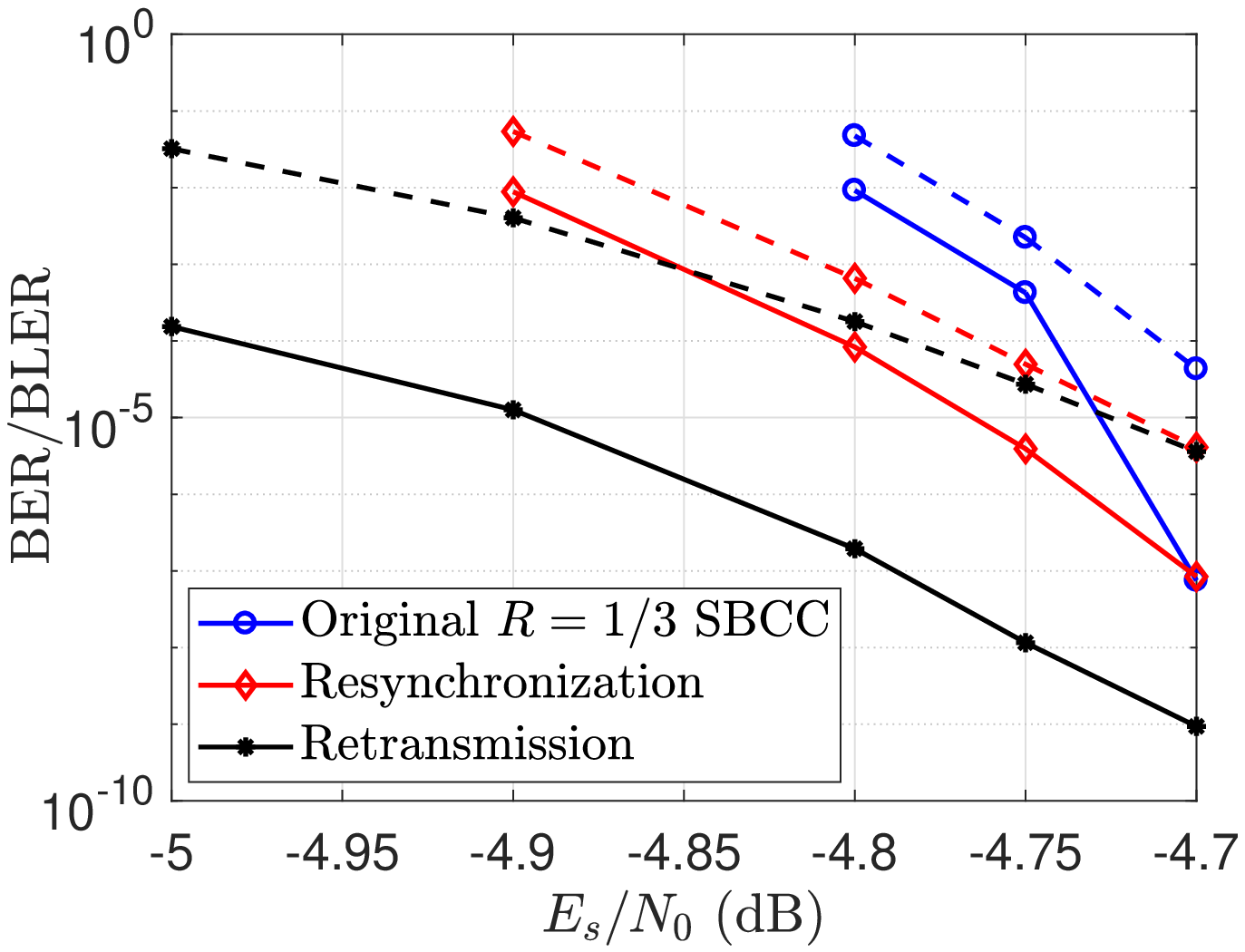}}
    \hspace{0.1cm}
    \subfigure[$T=500$, ${N'_r}=1$, and $w=6$.]{
            \label{fig:P500W6ExtRysTran}
            \includegraphics[width= 0.47 \textwidth]{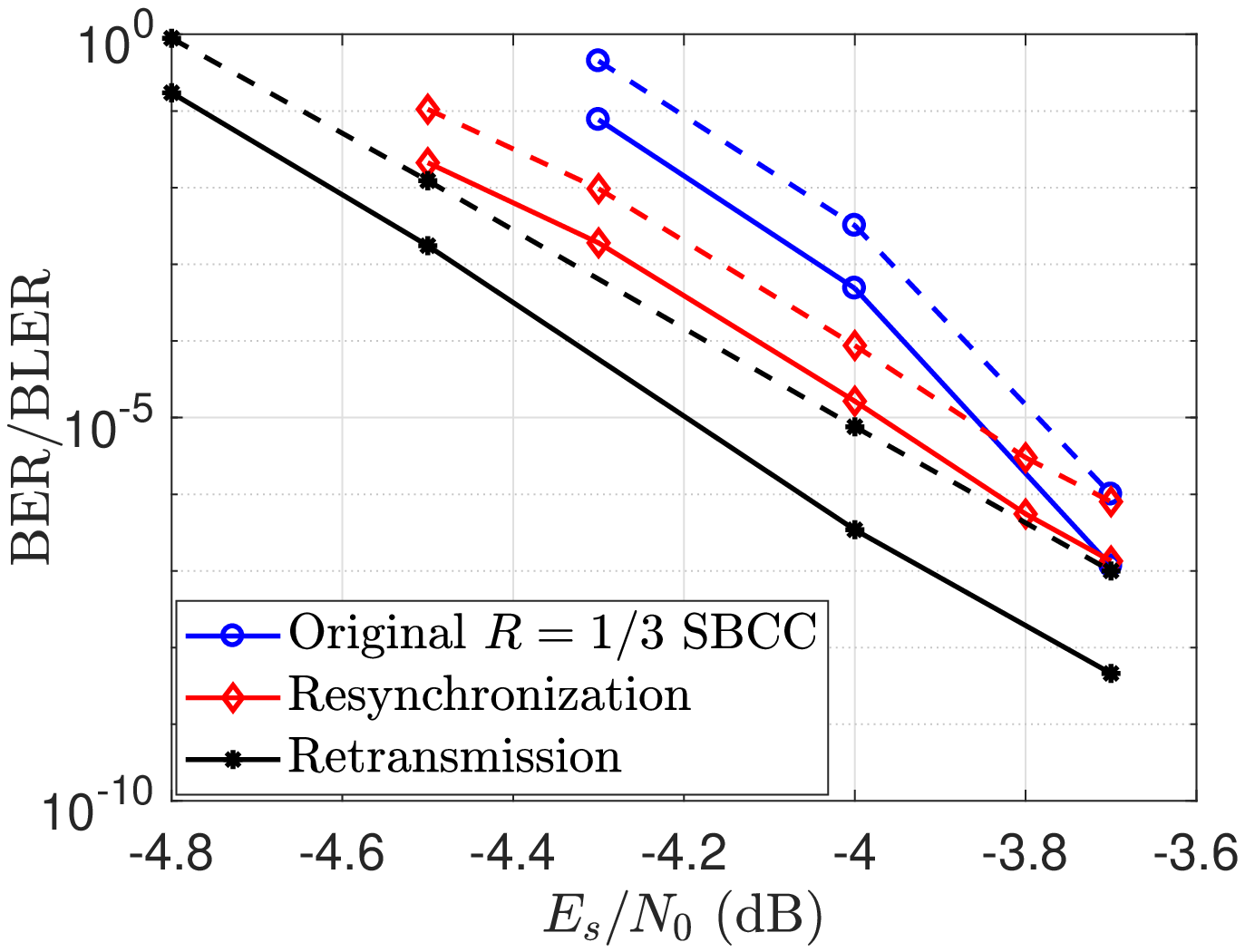}}
  \caption{BER (solid curves) and BLER (dashed curves) comparison of a rate $R=1/3$ SBCC with both resynchronization and retransmission.}\label{fig:TaoWin}
  \vspace{-4mm}
  \end{figure}

\emph{Example \myBlue{5}}: We first consider the rate $R=1/3$ blockwise SBCC of Example 1 with $T=8000$, $N'_r = 2$, and $w=3$. The BER/BLER performance with both resynchronization and retransmission is shown in \myBlue{Fig. \ref{fig:BER_ResetRateloss}.\footnote{In \myBlue{Figs. \ref{fig:BER_ResetRateloss} and \ref{fig:P500W6ExtRysTran}}, we plot the performance in terms of $E_s/N_0$ rather than $E_b/N_0$, since the average effective rate ${\tilde R}$ changes depending on the channel noise conditions. \myBlue{The chosen values of $N'_r$ were optimized empirically in both cases.}\label{fnExample4}}}
Compared to the $R=1/3$ blockwise SBCC of Example 1, resynchronization gains about two orders of magnitude and retransmission almost four orders of magnitude in BER, while the gains in BLER are about two orders of magnitude for resynchronization and slightly more for retransmission. We also see that the curves tend to merge as the SNR increases, as we have noted previously, i.e., the error propagation mitigation methods we propose help mainly in a narrow, but very important, range of SNRs, viz., the operating range in many applications. \hfill $\blacksquare$

\begin{figure}
    \centering
    \setlength{\subfigcapskip}{-0.3cm}
    \subfigure[\myBlue{$T=500$, $w=6$, original.}]{
        \label{fig:OriginalP500W6EsN4F}
        \includegraphics[width=  \textwidth]{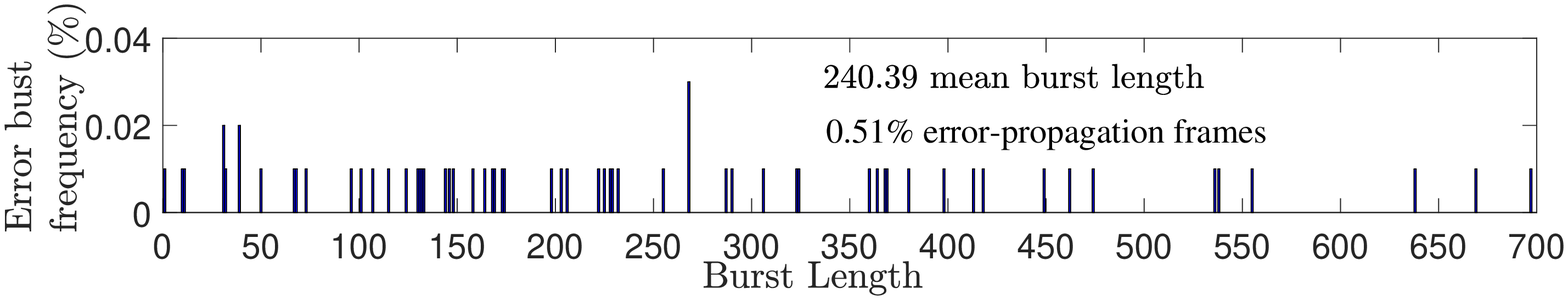}}
    \hspace{0.1cm}
    \subfigure[\myBlue{$T=500$, ${N_r}=1$, $w=6$, resynchronization.}]{
            \label{fig:P500W6EsN4FRysTran}
            \includegraphics[width=  \textwidth]{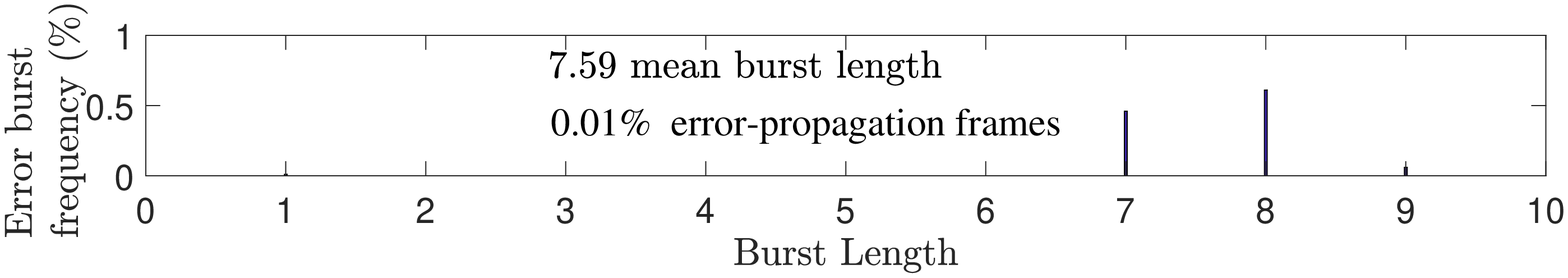}}
    \hspace{0.1cm}
    \subfigure[\myBlue{$T=500$, ${N'_r}=1$, $w=6$, retransmission.}]{
            \label{fig:P500W6EsN4FRetrans}
            \includegraphics[width= \textwidth]{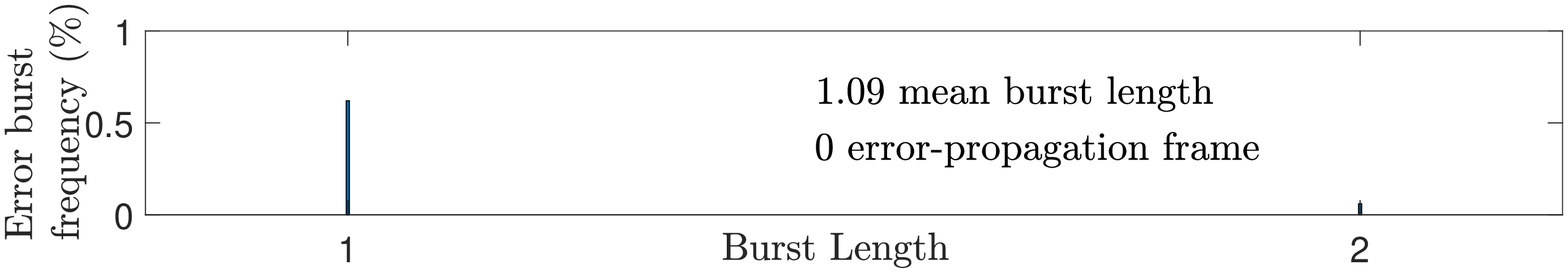}}
  \caption{\myBlue{The frequency of the error frames in a rate $R=1/3$ SBCC with and without resynchronization and retransmission for $T = 500$ at $E_s/N_0 = -4$ dB.}}\label{fig:P500W6EsN4F}
  \vspace{-8mm}
  \end{figure}

\emph{Example \myBlue{6}}: We next consider the rate $R=1/3$ blockwise SBCC of Example 1 with $T = 500$, $N'_r = 1$, and $w=6$. The BER/BLER performance with both resynchronization and retransmission is shown in \myBlue{Fig. \ref{fig:P500W6ExtRysTran}}.
Compared to the $R=1/3$ blockwise SBCC of Example 1, resynchronization again gains about two orders of magnitude and retransmission almost four orders of magnitude in BER, while the gains in BLER are almost two and three orders of magnitude, respectively. \myBlue{The frequencies of the burst-error frames and error-propagation frames, along with the mean burst length, are also given in Fig. \ref{fig:P500W6EsN4F}, which shows that both retransmission and resynchronization provide significant performance improvements, but that retransmission is best. }\hfill $\blacksquare$

\subsection{Window Extension plus Retransmission}
Retransmission can also be combined with window extension. Similar to the case of window extension with resynchronization, the decoder tries window extension until $w=w_{\max}$ and \myBlue{(\ref{eq:2})} is still satisfied, and then it checks if the retransmission condition \myBlue{\eqref{eq:Reset}} is satisfied. Algorithm \ref{Alg:WinExtRetran_Alg} in the appendix illustrates the details.

The BER/BLER/FER performance of the rate $R=1/3$ blockwise SBCC of Example 1 employing window extension plus retransmission is shown in \myBlue{Fig. \ref{fig:P500W6ExtReTran}} for $T=500$, ${N'_r} = 2$, $w_{\rm{init}}=6$, $w_{\max}=12$, $\tau=2$, and $\theta = 10$. We see that, compared to the rate $R=1/3$ blockwise SBCC of Example 1, the SBCC with window extension and retransmission gains close to one order of magnitude in FER, more than three orders of magnitude in BLER, and four orders of magnitude in BER in the SNR operating range of interest, exceeding the gains obtained with window extension plus resynchronization shown in \myBlue{Fig. \ref{fig:BER_WinExtReset}}. This confirms the fact that the retransmission eliminates some of the error blocks that remain following resynchronization. Also, comparing \myBlue{Fig. \ref{fig:P500W6ExtReTran} to Fig. \ref{fig:P500W6ExtRysTran}} illustrates the advantage of combining window extension and retransmission.


\section{Early Stopping Rule}

The decoding complexity of SBCCs with sliding window decoding depends mainly on the number of horizontal iterations. Therefore, in order to minimize unnecessary horizontal iterations, we introduce a soft BER stopping rule, which was first proposed for spatially coupled LDPC codes in \myBlue{\cite{Najeeb2016TCOM}.\footnote{\myBlue{Other stopping rules, such as the cross-entropy rule from \cite{Zhu2017TCOM}, could be employed here. However, since the LLR magnitudes must be used  anyway in the mitigation methods, it is easy to use them also to compute the soft BER estimates.}}}
Every time a horizontal iteration finishes, the average \textit{estimated bit error rate} $\text{BER}_{\text{est}}$ of the target bits in the current window is obtained using the following steps:
\begin{itemize}
  \item Calculate the decision LLR (the sum of the channel LLR, the prior LLR, and the extrinsic LLR) ${\ell}^{j}$ of every information bit in the target block, $j=0,1,\ldots, T-1$;
  \item Compute the average estimated BER of the target information bits as $${\rm{BER}_{\rm{est}}} = \frac{1}{T}\sum\limits_{j = 0}^{T - 1} {1.0/\left( {1.0 + \exp \left( {\left| {{\ell}^{j}} \right|} \right)} \right)}.$$
  \item If the average estimated BER of the target bits satisfies \myBlue{${{\rm{BER}}_{\rm{est}}} \le \gamma$,}
  decoding is stopped and a decision on the target symbols in the current window is made, where $\gamma$ is a predefined \emph{threshold} value.
\end{itemize}

\begin{figure}[htbp]
\centering
\begin{minipage}[t]{0.485\textwidth}
\centering
\includegraphics[width=8cm]{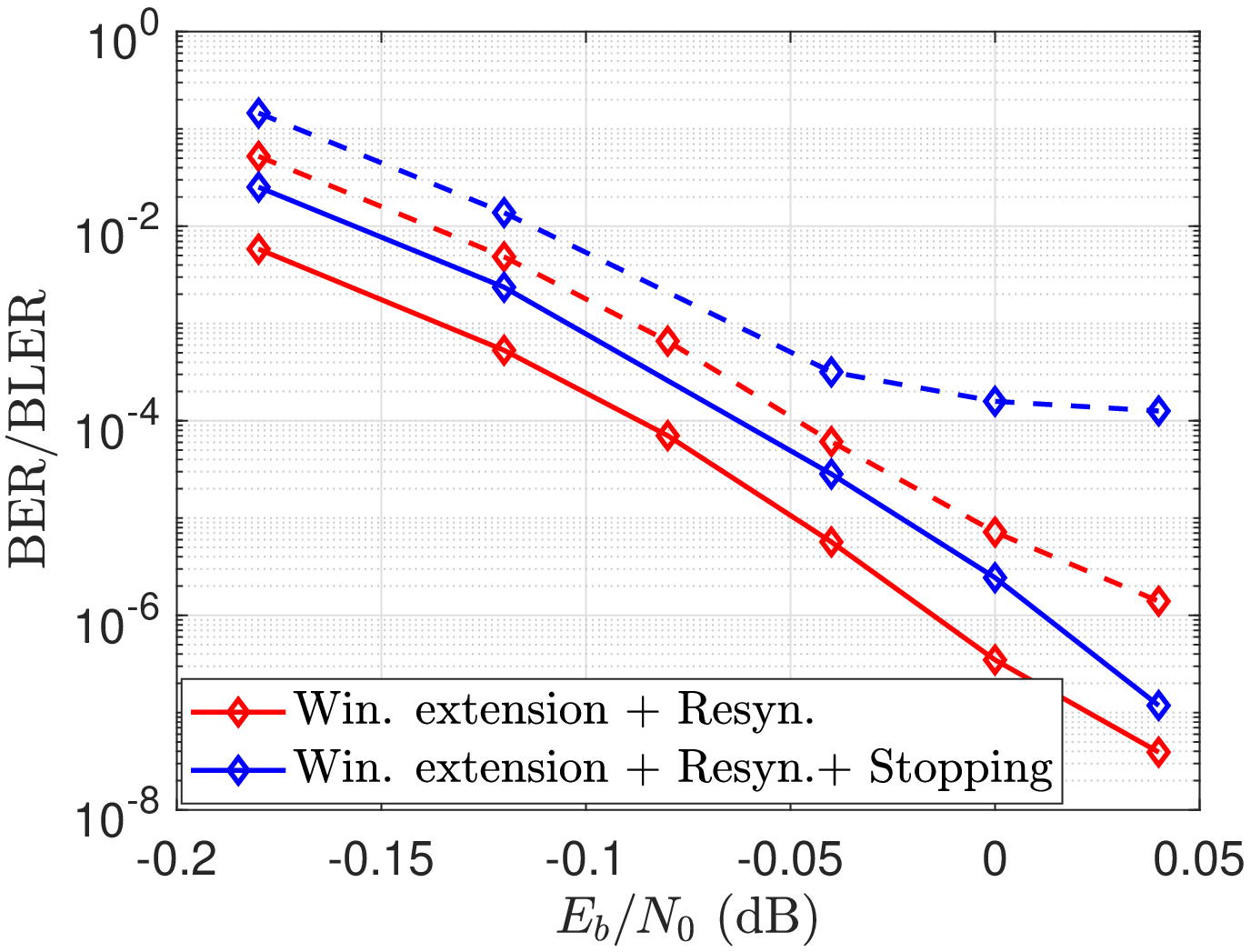}
\vspace{-7mm}
\caption{BER (solid curves) and BLER (dashed curves) comparison of a rate $R=1/3$ SBCC with window extension and resynchronization, with and without the soft BER stopping rule.}
\label{fig:BER_WinExtResetStop}
\end{minipage}
\hspace{0.1cm}
\begin{minipage}[t]{0.485\textwidth}
\centering
\includegraphics[width=8cm]{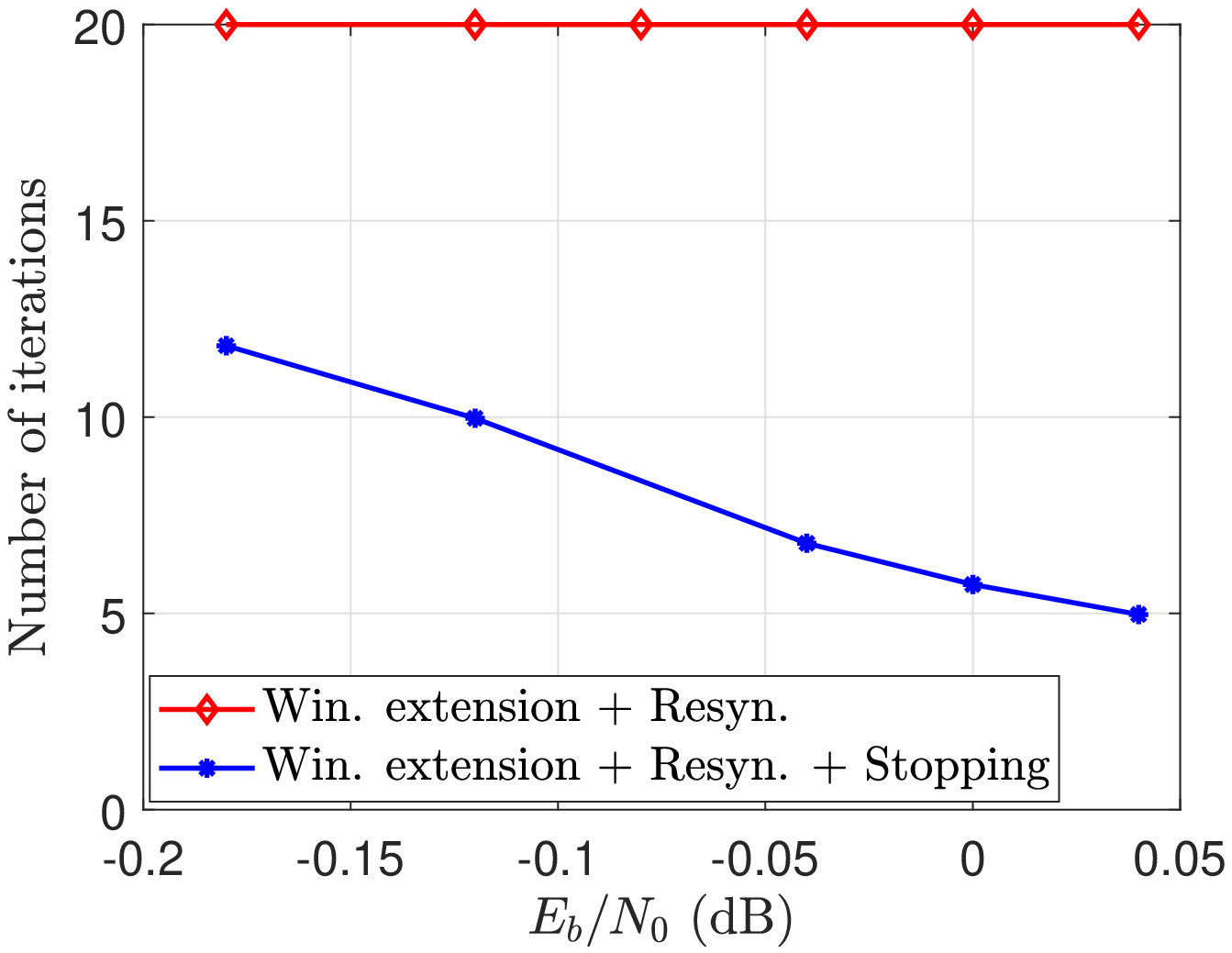}
\vspace{-7mm}
\caption{Number of horizontal iterations of a rate $R=1/3$ SBCC with window extension and resynchronization, with and without the soft BER stopping rule.}
\label{fig:ITER_WinExtResetStop}
\end{minipage}
\vspace{-10mm}
\end{figure}

Note that window extension, resynchronization, and the soft BER stopping rule can operate together in a sliding window decoder.
We now give an example to illustrate the tradeoffs between performance and computational complexity when these error propagation mitigation schemes are combined with the soft BER stopping rule. \myBlue{Fig. \ref{fig:BER_WinExtResetStop}} shows the performance of the rate $R=1/3$ blockwise SBCC of Example 1 with window extension, resynchronization, and the soft BER stopping rule for the same simulation parameters used in \myBlue{Fig. \ref{fig:BER_WinExtReset}} and $\gamma = 5 \times 10^{-8}$. We see that using the stopping rule degrades the BER performance only slightly, but the BLER performance is negatively affected in the high SNR region.\myBlue{\footnote{The BLER loss at high SNR can be reduced by using a smaller $\gamma$, at a cost of some increased decoding complexity, since a smaller $\tau$ results in a lower probability that a block will contain some bit errors.}} The average number of horizontal iterations per block is shown in Fig. \ref{fig:ITER_WinExtResetStop}, where we see that the soft BER stopping rule greatly reduces the required number of horizontal iterations, especially in the high SNR region.

\section{Conclusion}
In this paper we investigated the severe but infrequent error propagation problem associated with blockwise SBCCs and low latency sliding window decoding, which can have a catastrophic effect on performance for large frame lengths and continuous streaming operation.
We began by examining the causes of error propagation in sliding window decoding of SBCCs, noting that it is always accompanied by near zero average LLR magnitudes in the incorrectly decoded blocks. Based on this observation, a window extension algorithm, a resynchronization mechanism, and a retransmission strategy were proposed to mitigate the error propagation.
The FER, BLER, and BER of blockwise SBCCs with these three error propagation mitigation methods was shown to improve performance by up to four orders of magnitude in the SNR operating range of interest. Furthermore, a soft BER stopping rule was introduced and shown to significantly  reduce decoding complexity with only a slight effect on BER performance.

\ifCLASSOPTIONcaptionsoff
  \newpage
\fi

\newpage
\appendix
\vspace{-5mm}
\begin{algorithm}[H]
\caption{Window Extension Algorithm } \label{Alg:WinExt_Alg}
\begin{algorithmic}[1]
\STATE Assume that the block at time $t$ is the target block in a window decoder of size $w$ initialized with the channel LLRs of $w$ received blocks. Let $I_{\rm{count}}$ denote the current number of completed horizontal iterations, and set $I_{\rm{count}}=0$ and $w=w_{\rm{init}}$ initially, and let $\tau$, $\theta$, $w_{\rm{init}}$, and $w_{\max}$ be parameters.
\WHILE{$I_{count} < I_2$}
\STATE Perform vertical decoding and horizontal decoding;
\STATE Every time a horizontal iteration is finished,
\STATE $I_{\rm{count}} ++$;
\IF {$I_{\rm{count}} == I_2$}
  \STATE Calculate $\bar {\ell}^{\left( {i,j} \right)}$ according to \eqref{eq:ld}.
  \IF{\eqref{eq:2} is satisfied}
    \IF{$w < w_{\max}$}
     \STATE The decoder accepts one new block from the channel. The target block is still the block at time $t$, the new block is at time $t+w$, and the window size is set to $w=w+1$.
     \STATE $I_{\rm{count}} = 0$.
     \STATE For the $w$ blocks in the window, initialize the decoder with the channel LLRs, and reset the extrinsic information to \textbf{0}.
    \ENDIF
  \ENDIF
\ENDIF
\ENDWHILE
\STATE Decode the target block, set current window size $w=w_{\rm{init}}$, and shift the window.
\end{algorithmic}
\end{algorithm}

\begin{algorithm}[h]
\caption{Resynchronization Algorithm} \label{Alg:Reset_Alg}
\begin{algorithmic}[1]
  \STATE Assume that the block at time $t$ is the target block in a window decoder of size $w$ initialized with the channel LLRs of $w$ received blocks. Let $I_{\rm{count}}$ and $R_{\rm{count}}$ denote the current number of horizontal iterations and the counter for the average absolute LLRs of the target blocks, respectively, set $I_{\rm{count}}=0$ and $R_{\rm{count}}=0$ initially, and let $\theta$ and $N_r$ be parameters.
\WHILE{$I_{\rm{count}} < I_2$}\label{Alg2_start}
  \STATE Perform vertical decoding and horizontal decoding;
  \STATE $I_{\rm{count}} ++$;
\ENDWHILE
~~\STATE Calculate the average absolute LLRs $\bar {\ell}^{\left( t,I_2 \right)}$ of the target block using \eqref{eq:ld},
\IF {\eqref{eq:Reset} is satisfied}
  \STATE $R_{\rm{count}}++$.
\ELSE
  \STATE $R_{\rm{count}}=0$.
\ENDIF
\IF {$R_{\rm{count}} == N_r$}
  \STATE Resynchronize the encoder and decoder: 1) the initial register state of each component convolutional encoder is set to ``0''; 2) the parity input sequence of each component encoder is set to ``0'', the other input sequence is the new information block; 3)set $I_{\rm{count}}=0$ and $R_{\rm{count}}=0$;  4) the decoder makes decisions based on the current LLRs for all blocks in the window and restarts decoding once $w$ new blocks are received.
\STATE Go to step \ref{Alg2_start}.
\ELSE
  \STATE Decode the target block and shift the window.
\ENDIF

\end{algorithmic}
\end{algorithm}

\begin{algorithm}[h]
\caption{Window Extension plus Resynchronization Algorithm } \label{Alg:WinReset_Alg}
\begin{algorithmic}[1]
\STATE Assume that the block at time $t$ is the target block in a window decoder of size $w$ initialized with the channel LLRs of $w$ received blocks. Let $I_{count}$ and $R_{count}$ denote the current number of horizontal iterations and the counter for the average absolute LLRs of the target blocks, respectively, set $I_{count}=0$, $R_{count}=0$, and $w=w_{\rm{init}}$ initially, and let $\tau$, $\theta$, $w_{\rm{init}}$,  $w_{\max}$, and $N_r$ be parameters.
\WHILE{$I_{count} < I_2$}
\STATE Perform vertical decoding and horizontal decoding. Every time a horizontal iteration is finished,
\STATE $I_{count} ++$;
\IF {$I_{count} == I_2$}
  \STATE Calculate $\bar {\ell}^{\left( {i,j} \right)}$ according to \eqref{eq:ld}.
  \IF{\eqref{eq:2} is satisfied}
    \IF{$w < w_{max}$}
     \STATE The decoder accepts one new block from the channel. The target block is still the block at time $t$, the new block is the block at time $t+w$, and the window size is set to $w=w+1$.
     \STATE $I_{count} = 0$.
     \STATE For the $w$ blocks in the window, initialize the decoder with the channel LLRs, and reset the extrinsic information to \textbf{0}.
    \ENDIF
  \ENDIF
\ENDIF
\ENDWHILE
\STATE Reset the window size to be $w_{\rm{init}}$: $w=w_{\rm{init}}$.
\algstore{myalg}
\end{algorithmic}
\end{algorithm}

\begin{algorithm}[h]
\begin{algorithmic}[1]
\algrestore{myalg}
\STATE Calculate the average absolute LLRs $\bar {\ell}^{\left( t,I_2 \right)}$ of the target block using \eqref{eq:ld},
\IF {\eqref{eq:Reset} is satisfied}
 \STATE $R_{count}++$.
\ELSE
 \STATE $R_{count}=0$.
\ENDIF
\IF {$R_{count} == N_r$}
\STATE Resynchronize the encoder and decoder: 1) the initial register state of each component convolutional encoder is set to ``0''; 2) the parity input sequence of each component encoder is set to ``0'', the other input sequence is the new information block; 3) set $I_{count}=0$ and $R_{count}=0$; 4) the decoder makes decisions based on the current LLRs for all blocks in the window and restarts decoding once $w$ new blocks are received.
\STATE Go to step 2.
\ELSE
\STATE Decode the target block and shift the window.
\ENDIF
\end{algorithmic}
\end{algorithm}

\begin{algorithm}[h]
\caption{Retransmission Algorithm } \label{Alg:Retran_Alg}
\begin{algorithmic}[1]
\STATE Assume that the block at time $t$ is the target block in a window decoder of size $w$ initialized with the channel LLRs of $w$ received blocks. Let $I_{count}$ and $R_{count}$ denote the current number of horizontal iterations and the counter for the average absolute LLRs of the target blocks, respectively, set $I_{count}=0$ and $R_{count}=0$ initially, and let $\theta$ and ${N^{'}}_r$ be parameters.
\WHILE{$I_{count} < I_2$}
\STATE Perform vertical decoding and horizontal decoding; $I_{count} ++$.
\ENDWHILE
\STATE Calculate the average absolute LLRs $\bar {\ell}^{\left( t,I_2 \right)}$ of the target block using \eqref{eq:ld}.
\IF {\eqref{eq:Reset} is satisfied}
 \STATE $R_{count}++$.
\ELSE ~~~$R_{count}=0$.
\ENDIF
\IF {$R_{count} == N^{'}_r$}
\STATE Initialize the encoder: 1) the initial register state of each component convolutional encoder is set to ``0''; 2) the parity input sequence of each component encoder is set to ``0'', the other input sequence is the new information block; 3) the $w-1$ remaining information blocks in the decoding window reenter the encoder in sequence. The corresponding encoded blocks are then retransmitted over the channel.
\STATE Initialize the decoder: 1) the decoder deletes all the LLRs (including the channel LLRs, the extrinsic LLRs, and the a priori LLRs) for the $w-1$ remaining blocks in the window; 2) the decoder is reset to the initial state; 3) when the $w-1$ retransmitted blocks, plus one ``new" block, are received, decoding restarts with the initialization of the corresponding channel LLRs of these $w$ recieved blocks; 5) Go to step 2.
\ELSE
\STATE Decode the target block and shift the window.
\ENDIF

\end{algorithmic}
\end{algorithm}


\begin{algorithm}[h]
\caption{Window Extension plus Retransmission Algorithm } \label{Alg:WinExtRetran_Alg}
\begin{algorithmic}[1]
\STATE Assume that the block at time $t$ is the target block in a window decoder of size $w$ initialized with the channel LLRs of $w$ received blocks. Let $I_{count}$ and $R_{count}$ denote the current number of horizontal iterations and the counter for the average absolute LLRs of the target blocks, respectively, set $I_{count}=0$, $R_{count}=0$, and $w=w_{\rm{init}}$ initially, and let $\tau$, $\theta$, $w_{\rm{init}}$,  $w_{\max}$, and ${N^{'}}_r$ be parameters.
\WHILE{$I_{count} < I_2$} \label{WinRetran}
\STATE Perform vertical decoding and horizontal decoding. Every time a horizontal iteration is finished,
\STATE $I_{count} ++$;
\IF {$I_{count} == I_2$}
  \STATE Calculate $\bar {\ell}^{\left( {i,j} \right)}$ according to \eqref{eq:ld}.
  \IF{\eqref{eq:2} is satisfied}
    \IF{$w < w_{max}$}
     \STATE The decoder accepts one new block from the channel. The target block is still the block at time $t$, the new block is the block at time $t+w$, and the window size is set to $w=w+1$.
     \STATE $I_{count} = 0$.
     \STATE For the $w$ blocks in the window, initialize the decoder with the channel LLRs, and reset the extrinsic information to \textbf{0}.
    \ENDIF
  \ENDIF
\ENDIF
\ENDWHILE
\STATE Reset the window size to be $w_{\rm{init}}$: $w=w_{\rm{init}}$.
\algstore{myalg}
\end{algorithmic}
\end{algorithm}

\begin{algorithm}[h]
\begin{algorithmic}[1]
\algrestore{myalg}
\STATE Calculate the average absolute LLRs $\bar {\ell}^{\left( t,I_2 \right)}$ of the target block using \eqref{eq:ld}.
\IF {\eqref{eq:Reset} is satisfied}
 \STATE $R_{count}++$.
\ELSE
 \STATE $R_{count}=0$.
\ENDIF
\IF {$R_{count} == N'_r$}
\STATE Initialize the encoder: 1) the initial register state of each component convolutional encoder is set to ``0''; 2) the parity input sequence of each component encoder is set to ``0'', the other input sequence is the new information block; 3) the $w-1$ remaining information blocks in the decoding window reenter the encoder in sequence. The corresponding encoded blocks are then retransmitted over the channel.
\STATE Initialize the decoder: 1) the decoder deletes all the LLRs (including the channel LLRs, the extrinsic LLRs, and the a priori LLRs) for the $w-1$ remaining blocks in the window; 2) the decoder is reset to the initial state; 3) when the $w-1$ retransmitted blocks, plus one ``new" block, are received, decoding restarts with the initialization of the corresponding channel LLRs of these $w$ recieved blocks.
\STATE Go to step \ref{WinRetran}.
\ELSE
\STATE Decode the target block and shift the window.
\ENDIF
\end{algorithmic}
\end{algorithm}

\end{document}